\documentclass[12pt, a4paper]{article}
\usepackage{indentfirst}
\usepackage{graphics}
\usepackage{epsf,epsfig}
\usepackage{authblk}

\def\vk{{\bf k}_{\perp}}

\def\gev{\,{\rm GeV}}
\begin{document}

\title{Study of transversity GPDs from pseudoscalar mesons production at EIC of China}
\author[1]{ S.V.Goloskokov\thanks{goloskkv@theor.jinr.ru}}
\author[2,3]{Ya-Ping Xie\thanks{xieyaping@impcas.ac.cn}}
\author[2,3,4]{Xurong Chen\thanks{xchen@impcas.ac.cn}} 
 \affil[1]{Bogoliubov Laboratory of Theoretical Physics, Joint Institute for Nuclear Research, Dubna 141980, Moscow region, Russia}
 \affil[2]{Institute of Modern Physics, Chinese Academy of Sciences, Lanzhou 730000, China}
 \affil[3]{University of Chinese Academy of Sciences, Beijing 100049, China}
 \affil[4]{Institute of Quantum Matter, South China Normal University, Guangzhou 510006, China}
\date{}
\maketitle
\begin{abstract}
    The exclusive $\eta$ and $\pi^0$ electroproduction is
studied in the handbag approach based on Generalized Parton
Distributions (GPDs) factorization. Predictions of $\pi^0$ and $\eta$ mesons
are calculated for future Electron-Ion Collider of China (EicC)
energy range using obtained cross sections we extract
information on the transversity GPDs contributions to these
processes.
\end{abstract}
\section{Leptoproduction of pseudoscalar mesons}
In this paper, we analyze pseudoscalar meson electroproduction
($\pi^0$, $\eta$) on the basis of handbag approach. Its essential
ingredients are the Generalized Parton Distributions (GPDs) that
were proposed in Refs \cite{muell,ji1,rad1} and provide an
extensive information on the hadron structure. GPDs are
complicated nonperturbative objects which depend on $x_{B}$ -the
momentum fraction of proton carried by parton, $\xi$- skewness and
$t$- momentum transfer. GPDs are connected in the forward limit
with Parton Distribution Functions (PDFs), they contain
information about hadron form factors and the parton angular
momentum \cite{ji2}. They give information on 3D structure of the
hadrons, see e.g. \cite{3d}. More details on GPDs can be found
e.g. in \cite{dvmp1,dvmp2,Diehl,Radyush}.

GPDs were proposed to investigate exclusive reactions such as
deeply virtual Compton scattering (DVCS) \cite{ji2,rad2,dvcs},
time-like Compton scattering (TCS) \cite{Berger:2001xd,
Pire:2008ea, Mueller:2012sma} and deeply virtual meson production
(DVMP)  \cite{dvmp1,dvmp2}. Such processes at large photon
vituality $Q^2$ can be factorized into the hard subprocess that
can be calculated perturbatively and the GPDs
\cite{ji2,rad2,dvcs}. Generally, this factorization was proved in
the leading-twist amplitude with longitudinally polarized photon.
This factorization formulae is valid up to power corrections of
the order $1/Q$ to the leading twist results which are unknown.

Study of exclusive meson electroproduction is one of the effective
way to access GPDs. Experimental study of $\pi^0$ production was
performed by CLAS \cite{clas} and COMPASS \cite{compass}. For
$\eta$ production CLAS results are available at \cite{cleta}.
These experimental data can be adopted to constrain the models of
GPDs.
 On the other hand, Electron-Ion Colliders (EICs) are the next
generation collider to study of nucleon structure.
USA and China both design to build the EICs in future \cite{eic,eicc,Chen:2020ijn}.
The GPDs property is one of the most important aims to investigate for the
EICs \cite{Chavez:2021koz}.

Theoretical investigation of DVMP in terms of GPDs is based on the
handbag approach where, as mentioned before, the amplitude is
factorized into the hard subprocess and GPDs \cite{ji1, rad1, ji2,
rad2} see Fig.~1. This amplitude has an ingredient the
non-perturbative meson Distribution Amplitudes, which probe the
two-quark component of the meson wave functions. One of the
popular way to construct GPDs is adopting so called Double
Distribution (DD) \cite{mus99} which construct $\xi$ dependencies
of GPDs and connect them with PDFs, modified by $t$- dependent
term. The handbag approach with DD form of GPDs was successfully
applied to the light vector mesons (VM) leptoproduction at high
photon virtualities $Q^2$ \cite{gk06} and the pseudoscalar mesons
(PM) leptoproduction \cite{gk09}.

In this work, we continue our previous study of $\pi^0$ production
\cite{gxc22} at the kinematics for EIC in China (EicC) based on
the handbag approach. As it was shown in \cite{gk09} the leading
twist longitudinal cross section $\sigma_L$ is rather small with
respect to the predominant contribution determined by transversely
polarized photons $\sigma_T$. This result was proved
experimentally by JLab Hall A collaboration \cite{halla}. The
transversity dominance $\sigma_T
\gg \sigma_L$ is confirmed in \cite{gxc22} at all EicC energy ranges for
$\pi^0$ production.

This paper is organized as follows. In section 2, we discuss the
 contributions to the meson production amplitudes from the
 transversity GPDs $H_T$, $\bar E_T$ \cite{gk11}.
 Within the handbag approach, the transversity GPDs
together with the twist-3 meson wave functions \cite{gk11}
contribute to the amplitudes with transversely polarized photons
which produce transverse cross sections $\sigma_T$. They give
essential contribution to the cross sections that are consistent
with experiment \cite{clas,compass}.

In section 3, we consider two models for transversity GPDs
that give results for the cross sections of the $\pi^0$ and $\eta$
leptoproduction that are consistent with experiment at CLAS and
COMPASS energies \cite{clas,compass,cleta}. Predictions  for
$\eta$ cross section at EicC energies are done. Later on we
extract information on the transversity GPDs contribution for
these reactions. We discuss possibility to perform $u,d$ flavor
separation for transversity GPDs $H_T$ and $\bar E_T$ using $\pi^0$ and
$\eta$ cross sections \cite{kr-conv, kubar}. Finally, We give some
 discussion and conclusions in section 4.

\section{Handbag approach. Properties of meson production amplitudes}
\begin{figure}
\begin{center}
    \includegraphics[width=10cm]{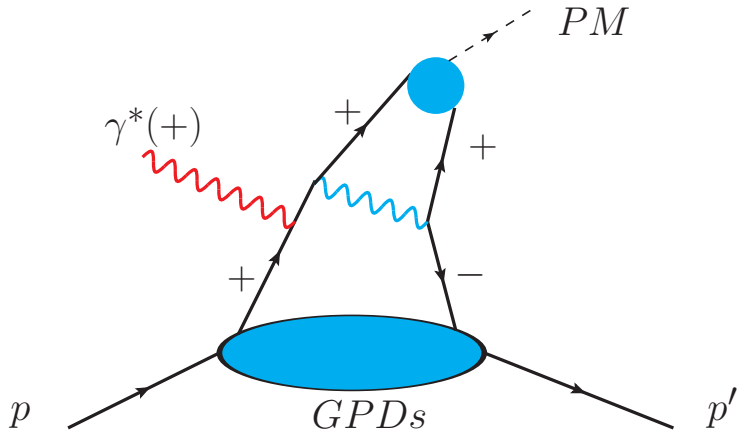}
\end{center}
    \caption{The
        handbag diagram for the meson electroproduction off proton. Parton
        helicities for transversity GPDs contribution are shown.}
    \label{kt_h}
\end{figure}

The process amplitude in the handbag approach is depicted in
Fig.~\ref{kt_h}. In handbag approach, the meson photoproduction
amplitude is factorized into a hard subprocess amplitude ${\cal
H}$ which is shown in the upper part of Fig.~\ref{kt_h} and GPDs
$F$ which includes
information on the hadron structure at sufficiently high $Q^2$. 
For the leading twist amplitude, with longitudinally polarized
photons, its factorization has been proved \cite{ji1,rad1}.

In what follows, we consider the twist-3 contributions from
transversity GPDs $H_T$ and $\bar {E}_T$ as well. Factorization
for these twist-3 amplitudes is an assumption now. However
factorization models give results which are consistent with
experiment  \cite{gk11}.

In handbag method, the subprocess amplitude is calculated employing
the modified perturbative approach (MPA) \cite{sterman}. The power
$k_\perp^2/Q^2$ correction is considered in the propagators of the
hard subprocess ${\cal H}$ together with the nonperturbative
$\vk$-dependent meson wave functions \cite{koerner}. The gluonic
corrections are regarded as the form of the Sudakov factors.
Resummation of the Sudakov factor can be done in the impact
parameter space \cite{sterman}.

The unpolarized $e p\to e(\pi^0, \eta)p$ cross section can be decomposed
into a number of partial cross sections which are expressed in
terms of the $\gamma^* p\to (\pi^0, \eta) p$ helicity amplitudes. They have
the following forms
\begin{eqnarray}\label{ds}
\frac{d\sigma_L}{dt} &=& \frac{1}{\kappa}
\big(\vert{M}_{0+,0+}\vert^2 +\vert{M}_{0-,0+}\vert^2\big)\,,\nonumber\\ [0.3em]
\frac{d\sigma_T}{dt} &=& \frac{1}{2 \kappa}(\vert{
    M}_{0-,++}\vert^2  +2 \vert{M}_{0+,++}\vert^2)\,, \nonumber\\[0.3em]
\frac{d\sigma_{LT}}{dt} &=& -\frac{1}{\sqrt{2} \kappa} {\rm
    Re}\Big[{M^*}_{0-,++}{M}_{0-,0+}\Big]
\,,\nonumber\\ [0.3em]
\frac{d\sigma_{TT}}{dt} &=& -\frac{1}{\kappa} \vert{M}_{0+,++}\vert^2\,.
\end{eqnarray}
Here $\kappa$ is the phase space factor, it reads
\begin{equation}\label{kap}
\kappa=16 \pi (W^2-m^2)\sqrt{\Lambda(W^2,-Q^2,m^2)}.
\end{equation}
 $\Lambda(x, y, z)$ is expressed as
 $\Lambda(x, y, z) = (x^2 + y^2 + z^2) - 2xy - 2xz - 2 yz$.
 $\sigma_{LT}$ is the interference contributions of the
 longitudinal and transverse amplitudes and $\sigma_{TT}$
contains transverse amplitudes only.

The leading twist amplitudes ${M}_{0-,0+} $  and ${M}_{0+,0+} $
are listed in our previous paper \cite{gxc22}.
The transversity amplitudes that are essential in our study can be
written in terms of convolutions as
\begin{eqnarray}\label{conv}
{M}_{0-,++}&=& \frac{e_0}{Q}\sqrt{1-\xi^2}\langle {H_T}\rangle,\nonumber\\
{M}_{0+,++}&=& -\frac{e_0}{Q}\frac{\sqrt{-t'}}{4m}\langle {\bar
    E_T}\rangle,
\end{eqnarray}
where $e_0 = \sqrt{4\pi \alpha}$ with $\alpha $ is
the electronic-magnetic coupling. The other variables are defined as
\begin{equation}
\xi=\frac{x_B}{2-x_B}(1+\frac{m_P^2}{Q^2}),\;\; t'=t-t_0,\;\; t_0=-\frac{4 m^2\xi^2}{1-\xi^2}.
\end{equation}
$x_B$ is the Bjorken variable which is given as $x_B = Q^2/(W^2 + Q^2 - m^2)$.
$m$ is the proton mass and $m_P$ is the pseudoscalar meson mass.

The GPDs $F(x,\xi,t)$ are calculated as the integration of the
double distributions function  \cite{mus99}
\begin{equation}
F(x,\xi,t) =  \int_{-1} ^{1}\, d\rho \int_{-1+|\rho|}
^{1-|\rho|}\, d\gamma \delta(\rho+ \xi \, \gamma - x) \,
f(\rho,t)\,\upsilon(\rho,\gamma,t).
\end{equation}
For the valence quark double distributions read as
\begin{equation}\label{ddf}
\upsilon(\rho,\gamma,t)=
\frac{3}{4}\,
\frac{[(1-|\rho|)^2-\gamma^2]}
{(1-|\rho|)^{3}}.
\end{equation}
The  $t$- dependence in PDFs $f$ is expressed as the Regge form
\begin{equation}\label{pdfpar}
f(\rho,t)= N\,e^{(b-\alpha' \ln{\rho})
    t}\rho^{-\alpha(0)}\,(1-\rho)^{\beta},
\end{equation}
and $\alpha(t)=\alpha(0)+\alpha' t$ is the corresponding Regge
trajectory factor. The parameters in Eq.~(\ref{pdfpar}) are fitted from
the known information about CTEQ6 PDF \cite{CTEQ6} e.g, or from the
nucleon form factor analysis \cite{pauli}. We consider $Q^2$
evolution of GPDs via evolution of PDF in Eq.~7, see
\cite{gk06}. This form of evolution is proper near the forward
limit. Generally in this work, the explicit form of GPDs evolution
is not so important because we work at very limited $Q^2$
interval.

It was found that for PM leptoproduction the contributions of
the transversity GPDs $H_T$ and $\bar {E}_T=2 \tilde H_T+E_T$ are
essential \cite{gk11}. It determines the amplitudes $M_{0-,++}$
and $M_{0+,++}$ respectively, see Eq.~(\ref{conv}). With the
handbag approach the transversity GPDs are accompanied by a
twist-3 PM wave functions in the hard amplitude ${\cal H}$
\cite{gk11} which is the same for both the ${M}_{0\pm,++}$
amplitudes in Eq.~(\ref{conv}). This property is demonstrated in
Fig.~\ref{kt_h}, where the parton helicities of the subprocess
amplitude ${\cal H}$ are presented. For corresponding transversity
convolutions we have forms:
\begin{equation}\label{ht}
\langle H_T\rangle =\int_{-1}^1 dx
{\cal H}_{0-,++}(x,...)\,H_T;\;
\langle \bar E_T\rangle =\int_{-1}^1 dx
{\cal H}_{0-,++}(x,...)\; \bar E_T.
\end{equation}
There is a parameter $\mu_P$ in twist-3 meson wave function that is large and enhanced by the
chiral condensate. In our calculation, we use $\mu_P$ = 2 \gev at scale of 2 \gev.

More details of leading twist polarized GPDs $\tilde H$ and
$\tilde E$ which contribute to the leading twist amplitudes with
longitudinally polarized virtual photons can be found in paper
\cite{gk09,gk11}. These amplitudes contribute to longitudinal
cross section $\sigma_L$ which is rather small with respect to
transversity contribution $\sigma_T$ for $\pi^0$ and $\eta$
production.

For additional information about transversity GPDs
parameterization see \cite{gk11} and \cite{gxc22}. The $\pi^0$
estimations at EicC are presented in our previous
paper \cite{gxc22}. Study of $\eta$ meson leptoproduction can be
performed within the handbag approach too, for details see \cite{gk11}.

\section{Model results for $\pi^0$ and $\eta$ leptoproduction and convolution extraction from the data}
We consider the transversity effects described in Eq.~(\ref{ht})
and take into account the leading twist contribution in
Eq.~(\ref{ds}). The amplitudes are transferred from program
 produced by PARTONS collaboration codes \cite{parton} which was
changed into Fortran employing results of GK model for GPDs
\cite{gk11}.

 In our previous paper \cite{gxc22}, two
models for transversity GPDs were analyzed. Model-1 was applied in
\cite{gk11} and described fine low energy CLAS data \cite{clas},
but gave results about two times larger with respect to COMPASS
data \cite{compass}. It was the reason to change GPDs parameters,
especially for $\bar E_T$ contribution that is important in
$\sigma_{T}$ and $\sigma_{TT}$ cross sections. Some changes were
done for $H_T$ as well. The parameters for new model labeled as
Model-2 are exhibited at Table.~1 \cite{krollpr}.

\begin{table*}[h]
    \renewcommand{\arraystretch}{1.4}
    \begin{center}
        \begin{tabular}{| c || c | c | c || c | c |}
            \hline GPD & $\alpha(0)$ & $\alpha^\prime [\gev^{-2}]$ & $b
            [\gev^{-2}]$ & $N^u$ &
            $N^d$ \\[0.2em]
            \hline
            $\bar{E}_T$& -0.1 & 0.45 & 0.67 & 29.23 & 21.61 \\[0.2em]
            $H_T$ & - & 0.45 & 0.04 & 0.68 & -0.186 \\[0.2em]\hline
        \end{tabular}
    \end{center}
    \caption{Regge parameters and normalizations of the GPDs at a
        scale of $2 \gev$ for Model-2.}
\end{table*}

\begin{figure}\label{comp}
    \epsfysize=65mm
    \centerline{\epsfbox{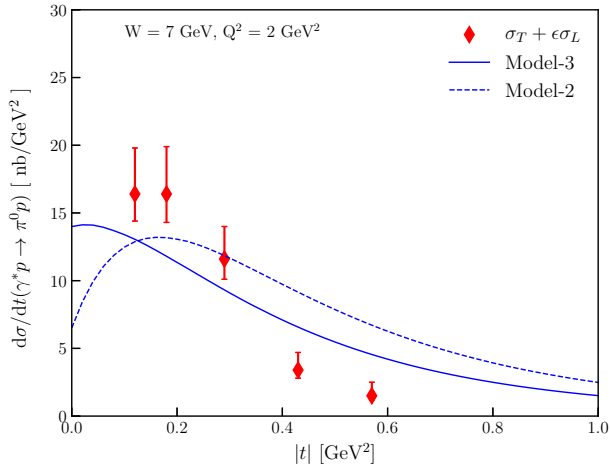}}
    \noindent\caption{Models results at COMPASS kinematics.
        Experimental data are taken from \cite{compass}, dashed line represents
        the results of Model-2 and solid curve indicates the prediction of Model-3}
\end{figure}

\begin{table*}[t]
    \renewcommand{\arraystretch}{1.4}
    \begin{center}
        \begin{tabular}{| c | c | c | c || c | c || c | c |}
            \hline GPD & $\alpha(0)$ & $\beta^u$& $\beta^d$&  $\alpha^\prime
            [\gev^{-2}]$ & $b [\gev^{-2}]$ & $N^u$ &
            $N^d$ \\[0.2em]
            \hline
            $\bar{E}_T$& -0.1 &4 & 5&  0.45 & 0.77 & 20.91 & 15.46 \\[0.2em]
            $H_T$ & - & -& -& 0.45 & 0.3 & 1.1 & -0.3 \\[0.2em]

            \hline
        \end{tabular}
    \end{center}
    \caption{Regge parameters and normalizations of the GPDs at a
        scale of $2\,\gev$. Model-3.}
\end{table*}

Results of this model are shown at COMPASS energies in Fig.~2 by
dashed lines. It can be seen that there are some discrepancy
between Model-2 results and COMPASS data \cite{compass} at large
$-t>0.3 \mbox{GeV}^2$. That was the reason to test in addition the
new Model-3 results for $\pi^0$ and $\eta$ leptoproduction. The
parameters for new Model-3 are listed at Table. 2
\cite{krollpr}. Note that in this model parameters are close to model I in
\cite{gxc22}, only parameters of $\bar{E}_T$ was changed.  It can
be seen from the $N$ parameters that Model-2 have larger
$\bar{E}_T$ and smaller $H_T$ values with respect to Model-3. In
Model-3, we have smaller $\bar{E}_T$ and larger $H_T$. Both models
describe well $\pi^0$ production at COMPASS. Model-3 gives better
 results for large $-t >0.3 \mbox{GeV}^2$, see Fig.~2.

\begin{figure}[h!]\label{cl-pi0}
    \begin{center}
    \begin{tabular}{cc}
            \includegraphics[width=7.5cm]{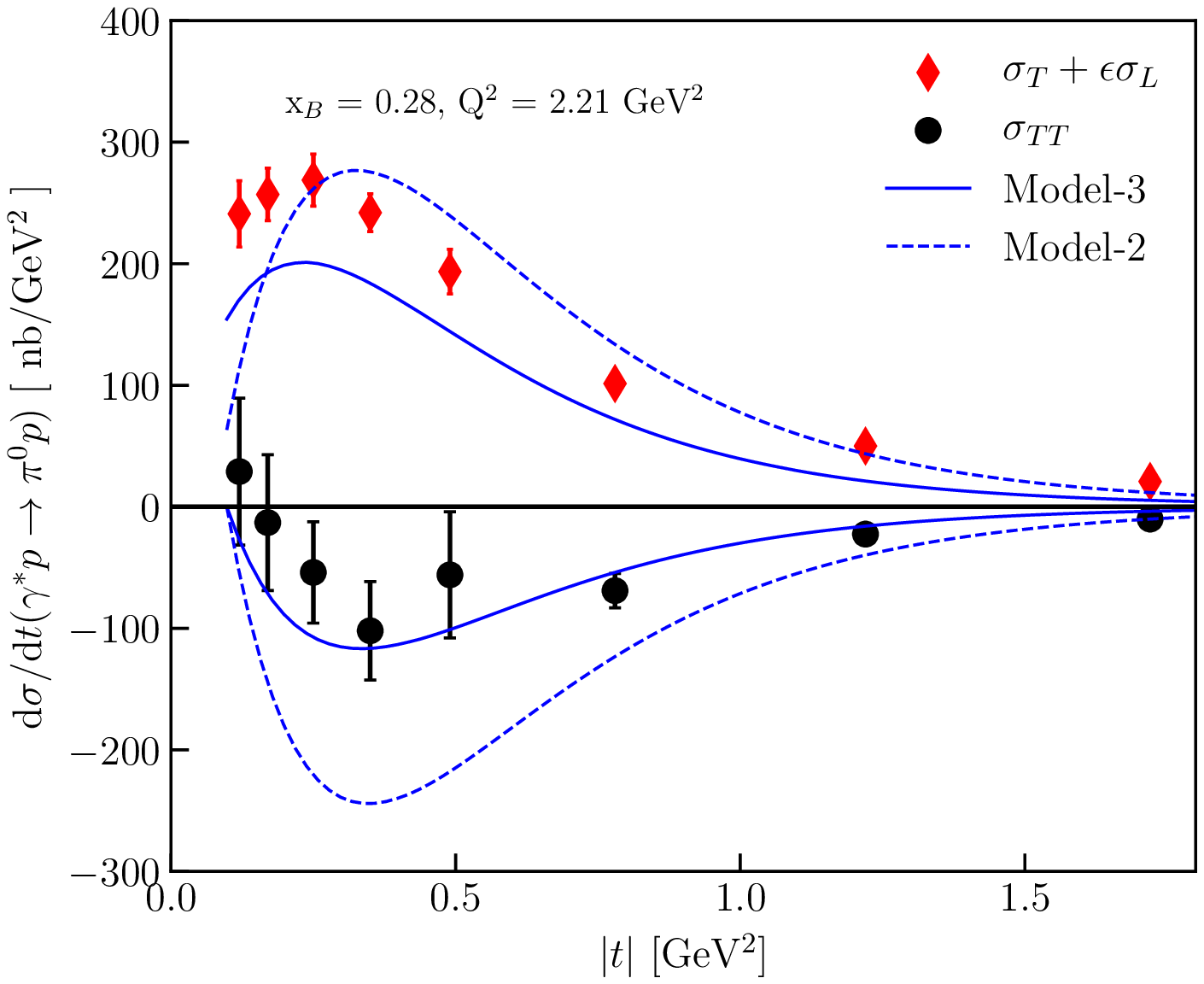}
            \includegraphics[width=7.5cm]{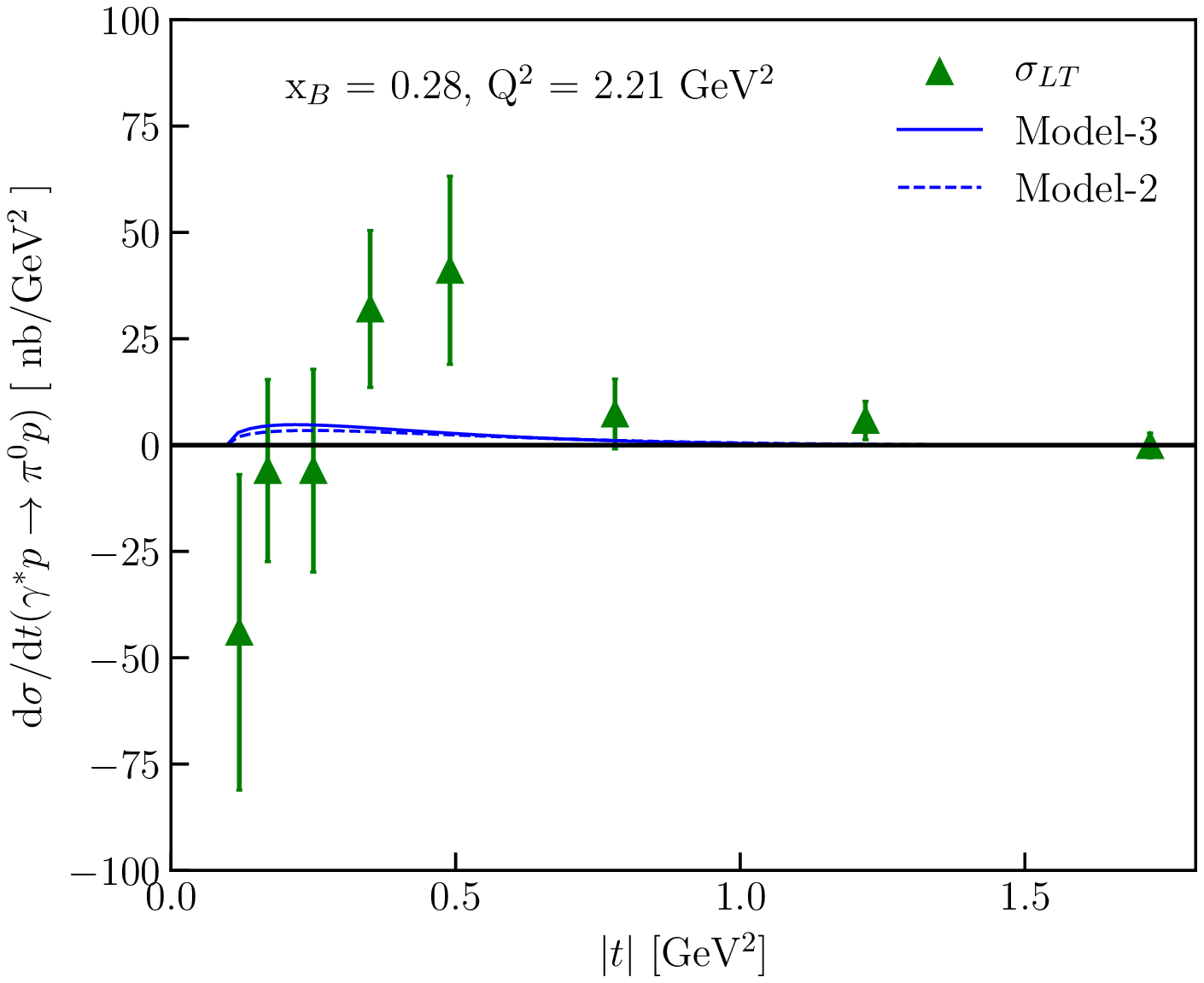}
\end{tabular}
    \end{center}
    \caption{Cross sections of $\pi^0$ production in the CLAS energy
        range together with the data \cite{clas}. Left graph is for $\sigma$
        and  $\sigma_{TT}$ while right graph is for $\sigma_{LT}$.
    }
\end{figure}

Model-2 and 3 results for $\pi^0$ production at CLAS energy are exhibited in
Fig.~\ref{cl-pi0}. It can be seen that both models are
in accordance with unseparated cross sections
$\sigma=\sigma_T+\epsilon \sigma_L$, where $\sigma_T$
predominated. At the same time, Model-3 gives closer results for
$\sigma_{TT}$ that is smaller with respect to Model-2. This
confirms mentioned before smaller value of $\bar{E}_T$ in the Model-3.
$\sigma_{LT}$ cross sections are shown as well.

Calculation of the amplitudes of  $\eta$ production is similar to
the $\pi^0$ case and based on the \cite{gk11} results where the
singlet-octet decomposition of $\eta$-state was used with
redefined decay constants.

The flavor factors for $\pi^0$ and $\eta$ production are appear in
combinations
\begin{eqnarray}\label{flf}
F_{\pi^0} =\frac{1}{\sqrt{2}} (e^u F^u-e^d F^d)=\frac{1}{3
    \sqrt{2}} (2 F^u+ F^d);\nonumber\\
F_{\eta} =\frac{1}{\sqrt{6}} (e^u F^u+e^d F^d)=\frac{1}{3
    \sqrt{6}} (2 F^u- F^d).
\end{eqnarray}
The $\eta$ factor is written without strange sea contribution
which is small and can be neglected.

From Table 1 and 2, it can be seen that $\bar {E}_T$ has the same signs for $u$ and
$d$ quarks but $H_T$ has the different signs, respectively. This
means that for $\pi^0$ case  $\bar E_T$ contributions for $u, d$
quarks are added but $H_T$ are subtracted. For $\eta$ production
we have opposite case: $H_T$ contributions are added but  $\bar
E_T$ compensated.

\begin{figure}[h!]\label{cl-eta}
    \begin{center}
        \begin{tabular}{cc}
            \includegraphics[width=7.5cm]{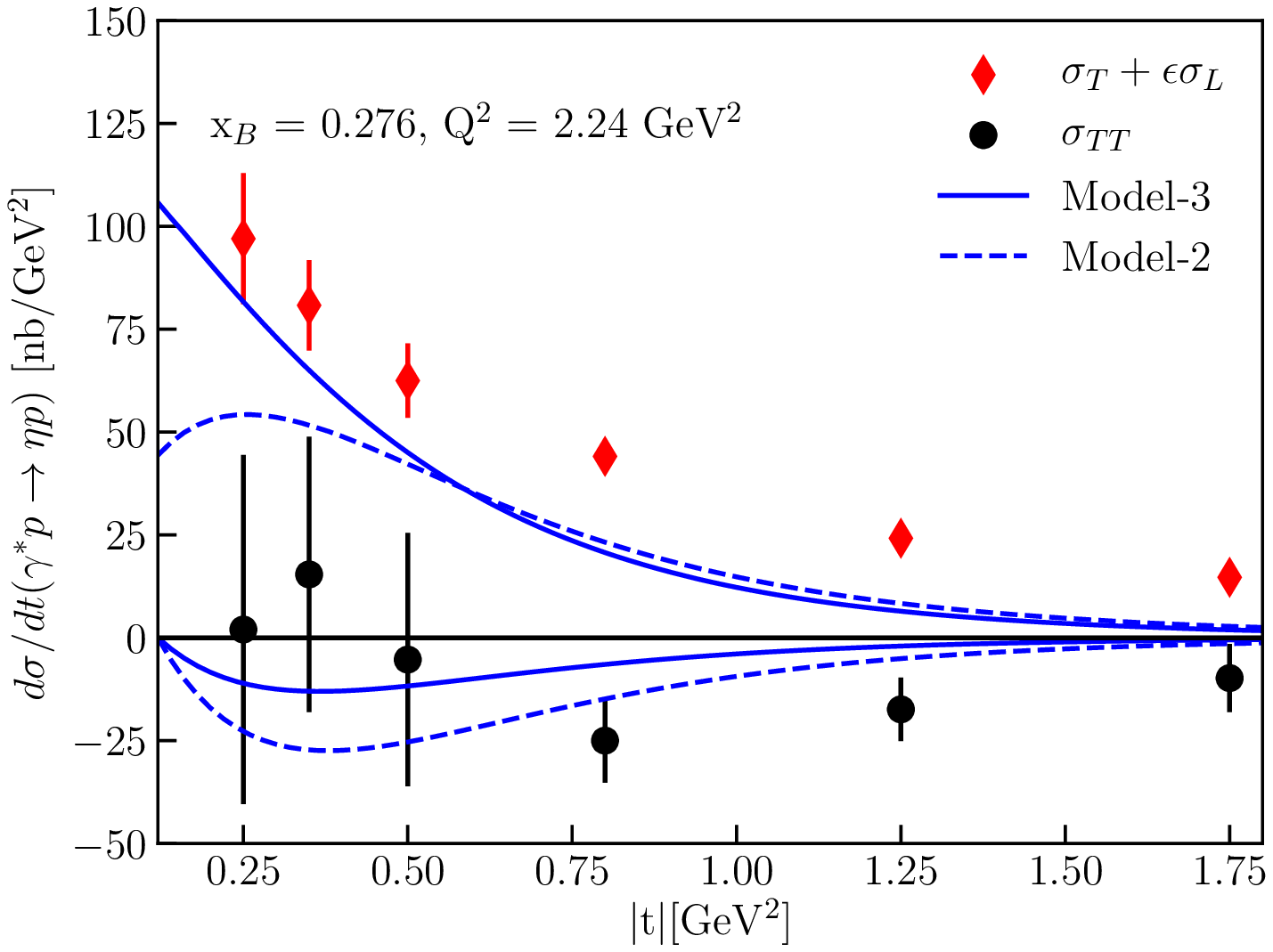}&
            \includegraphics[width=7.5cm]{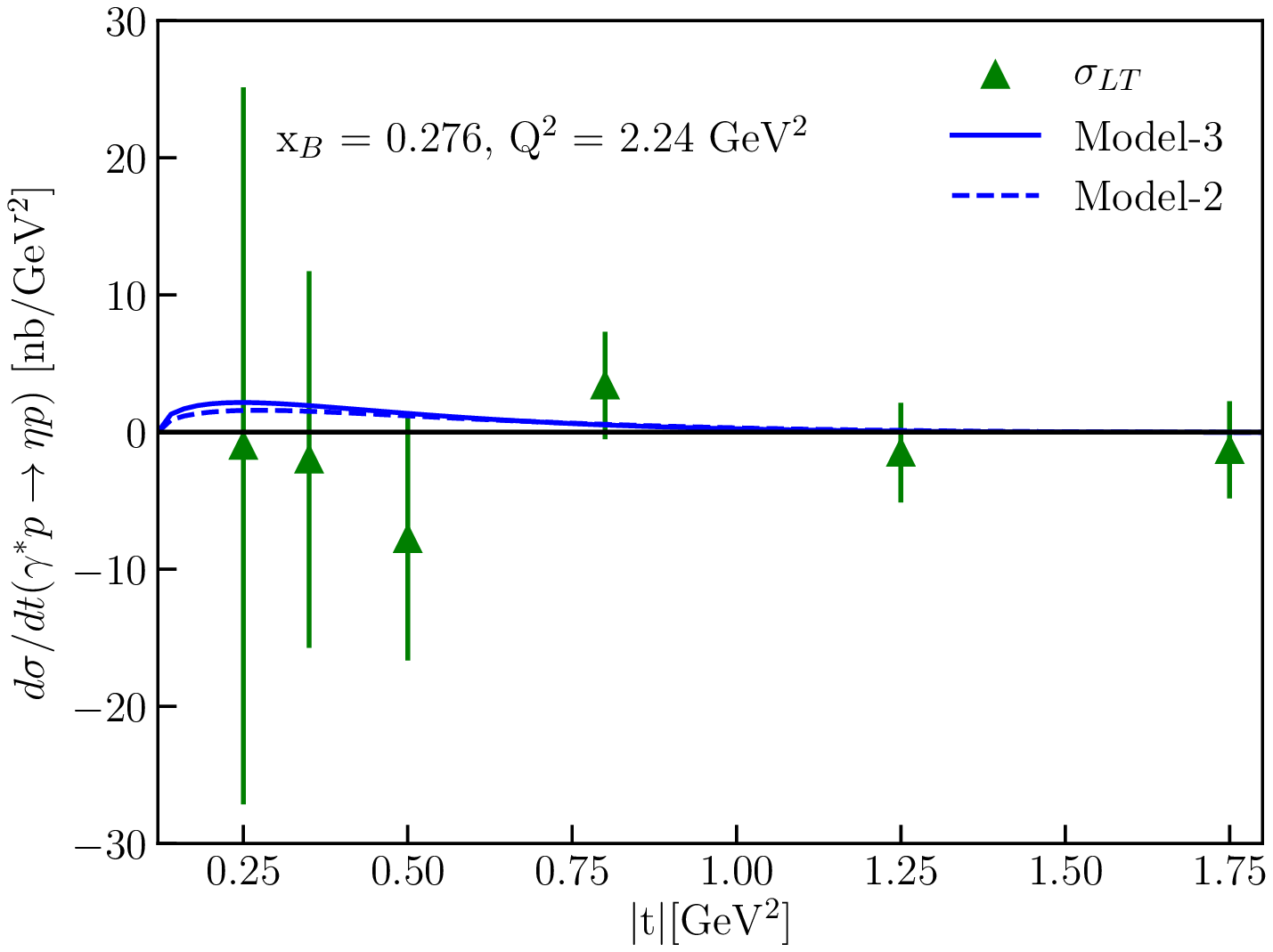}
        \end{tabular}
    \end{center}
    \caption{Cross section of $\eta$ production in the CLAS energy
        range together with the data \cite{cleta}. Left graph is for
        $\sigma$ and $\sigma_{TT}$ while right graph is for $\sigma_{LT}$.
    }
\end{figure}

\begin{figure}[h!]\label{eta78}
    \begin{center}
        \begin{tabular}{cc}
            \includegraphics[width=7.5cm]{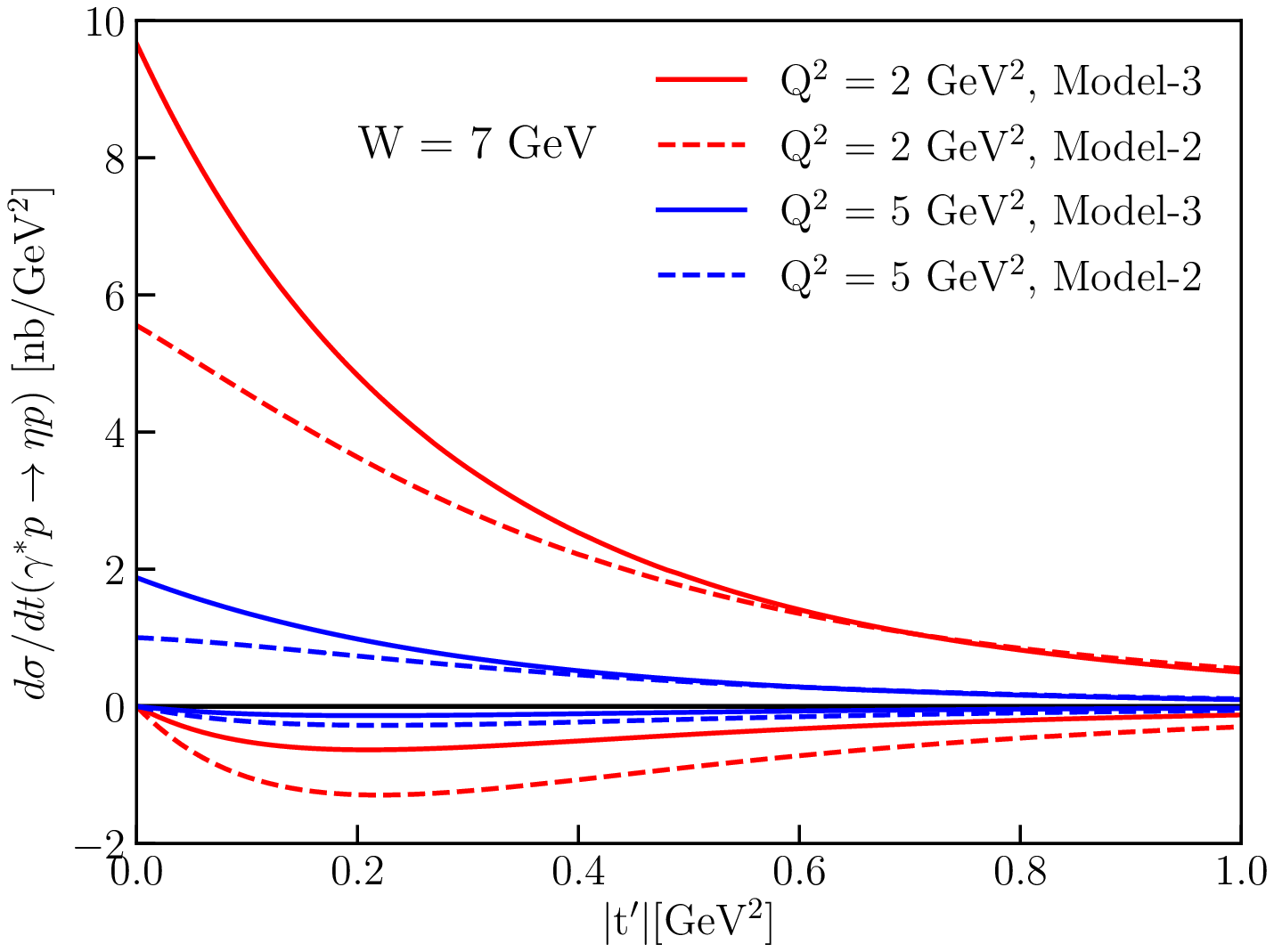}&
            \includegraphics[width=7.5cm]{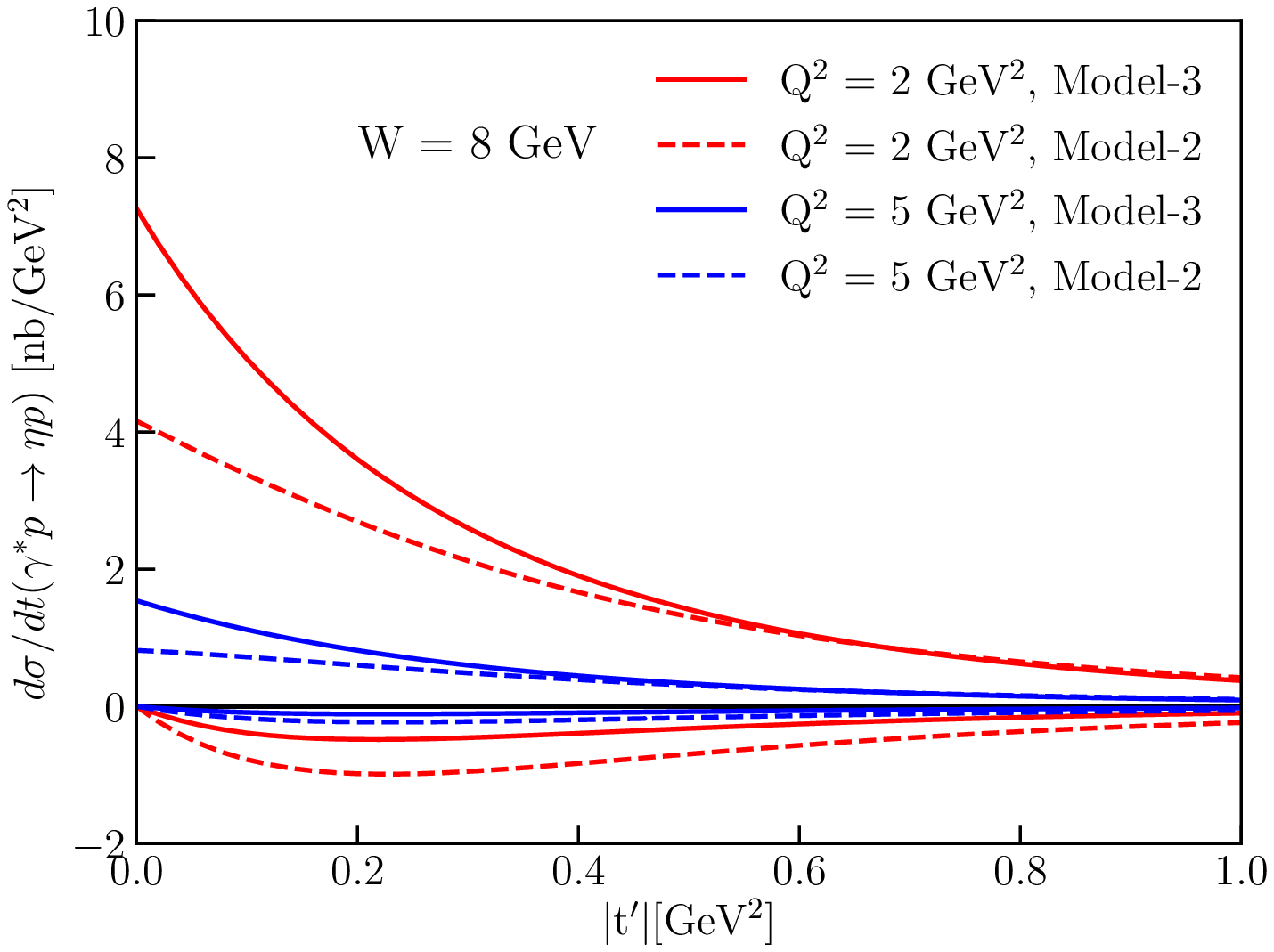}
        \end{tabular}
    \end{center}
    \caption{Cross sections of $\eta$ production at EicC energy.  Upper
        part of the figure presents $\sigma=\sigma_T+\epsilon\, \sigma_L$
        and down part- $\sigma_{TT}$ as in Fig.~4.
    }
\end{figure}

\begin{figure}[h!]\label{eta1216}
    \begin{center}
        \begin{tabular}{cc}
            \includegraphics[width=7.5cm]{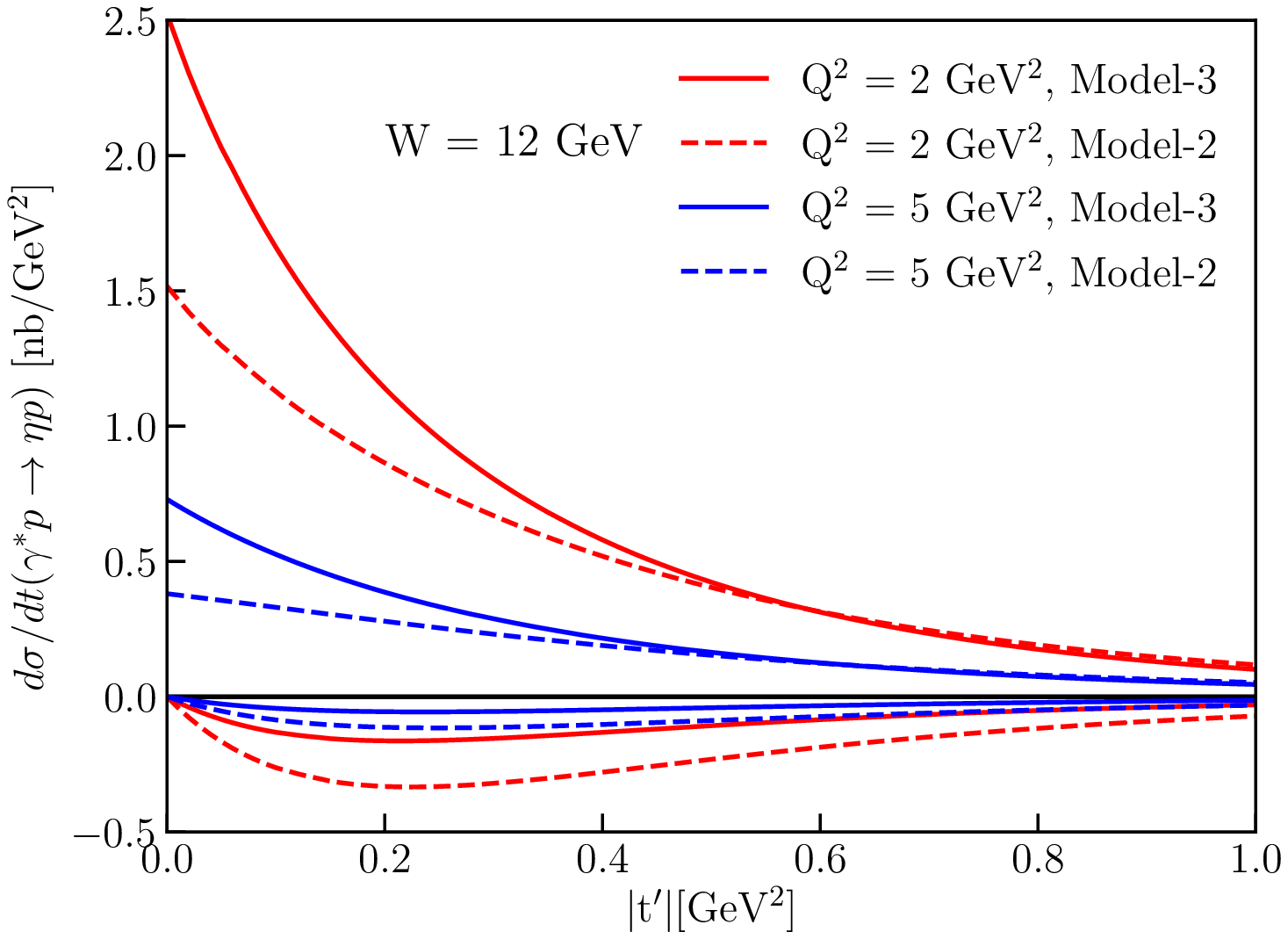}&
            \includegraphics[width=7.5cm]{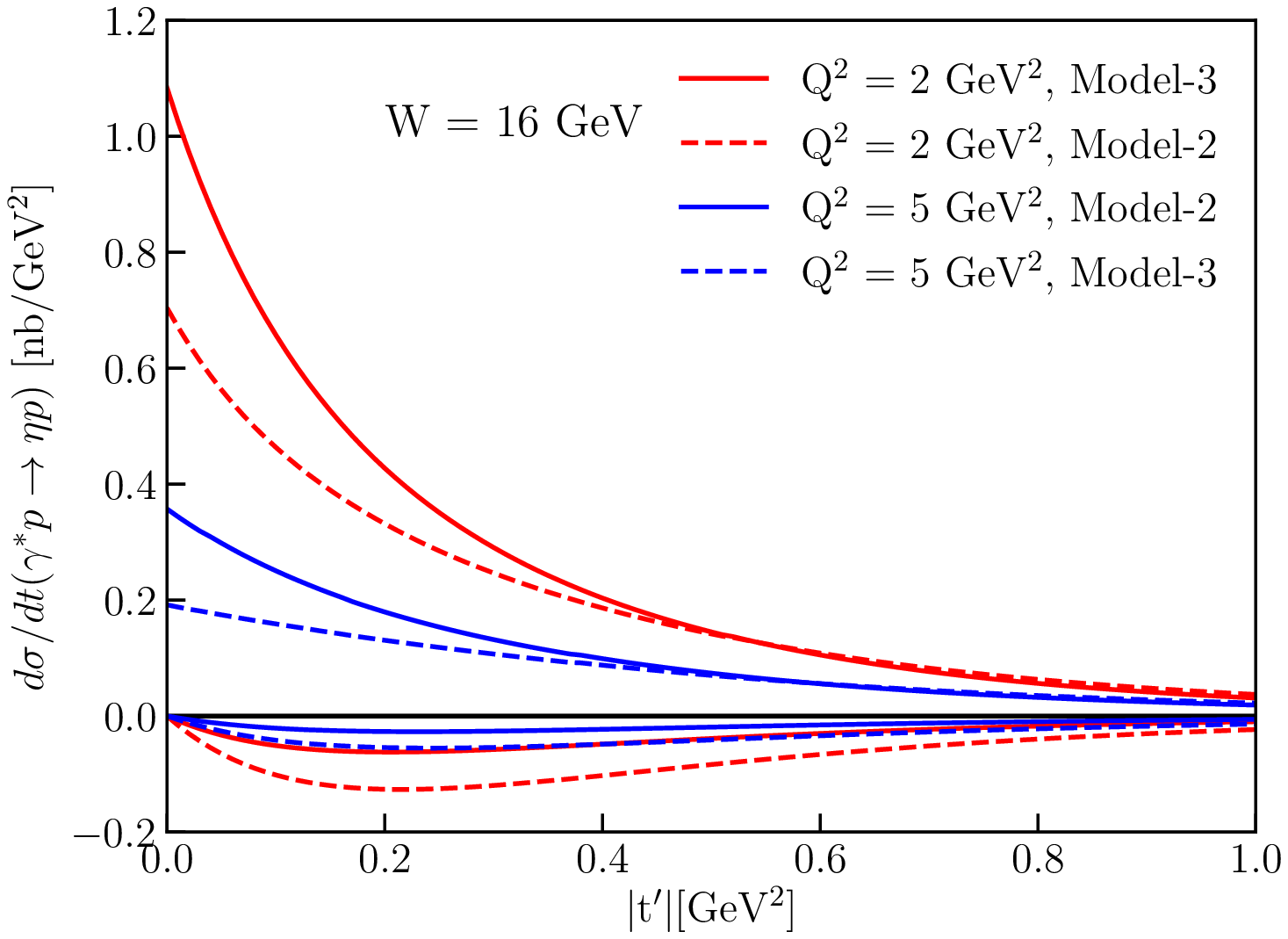}
        \end{tabular}
    \end{center}
    \caption{Cross section of $\eta$ production at EicC energy. The labels are
        same as in Fig.~5.
    }
\end{figure}
Thus we have $\bar E_T$ enhancement for $\pi^0$
case. For $\eta$ production $H_T$ is increasing. Therefore, $\pi^0$
process is more sensitive to $\bar E_T$ effects but for $\eta$
production $H_T$ influences are more visible.

Models results for $\eta$ production at CLAS energy \cite{cleta}
are depicted in Fig.~4. It can be seen that Model-3 with
larger $H_T$ contribution describes experimental data better at
small momentum transfer. Model-2 with smaller $H_T$ produces
essential dip in the cross section that is not observed at
experiment. Cross sections $\sigma_{TT}$ and $\sigma_{LT}$ are
described properly for both models.

\begin{figure}[h!]\label{conv-cl}
    \begin{center}
        \begin{tabular}{cc}
            \includegraphics[width=7.5cm]{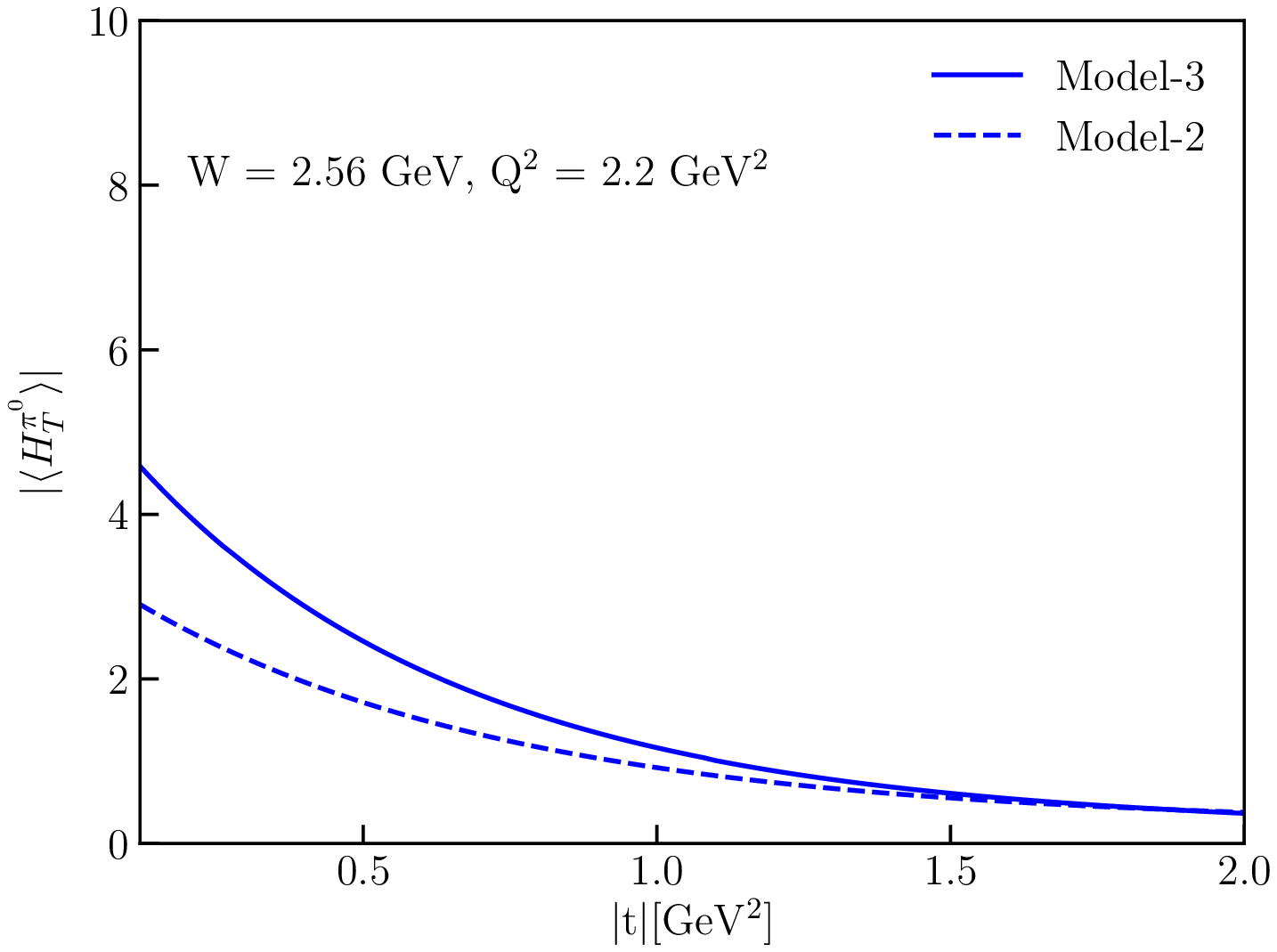}&
            \includegraphics[width=7.5cm]{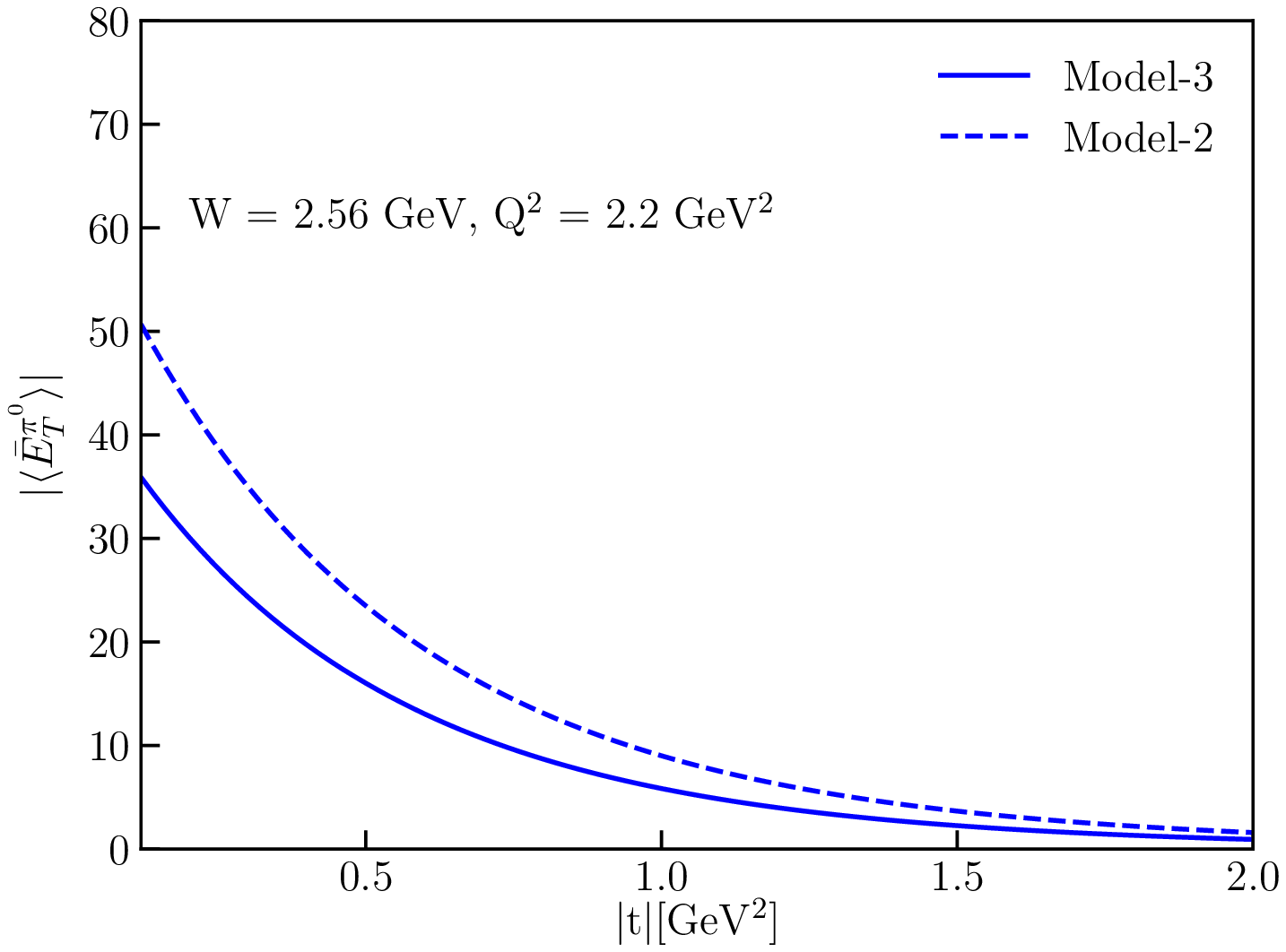}\\
            \includegraphics[width=7.5cm]{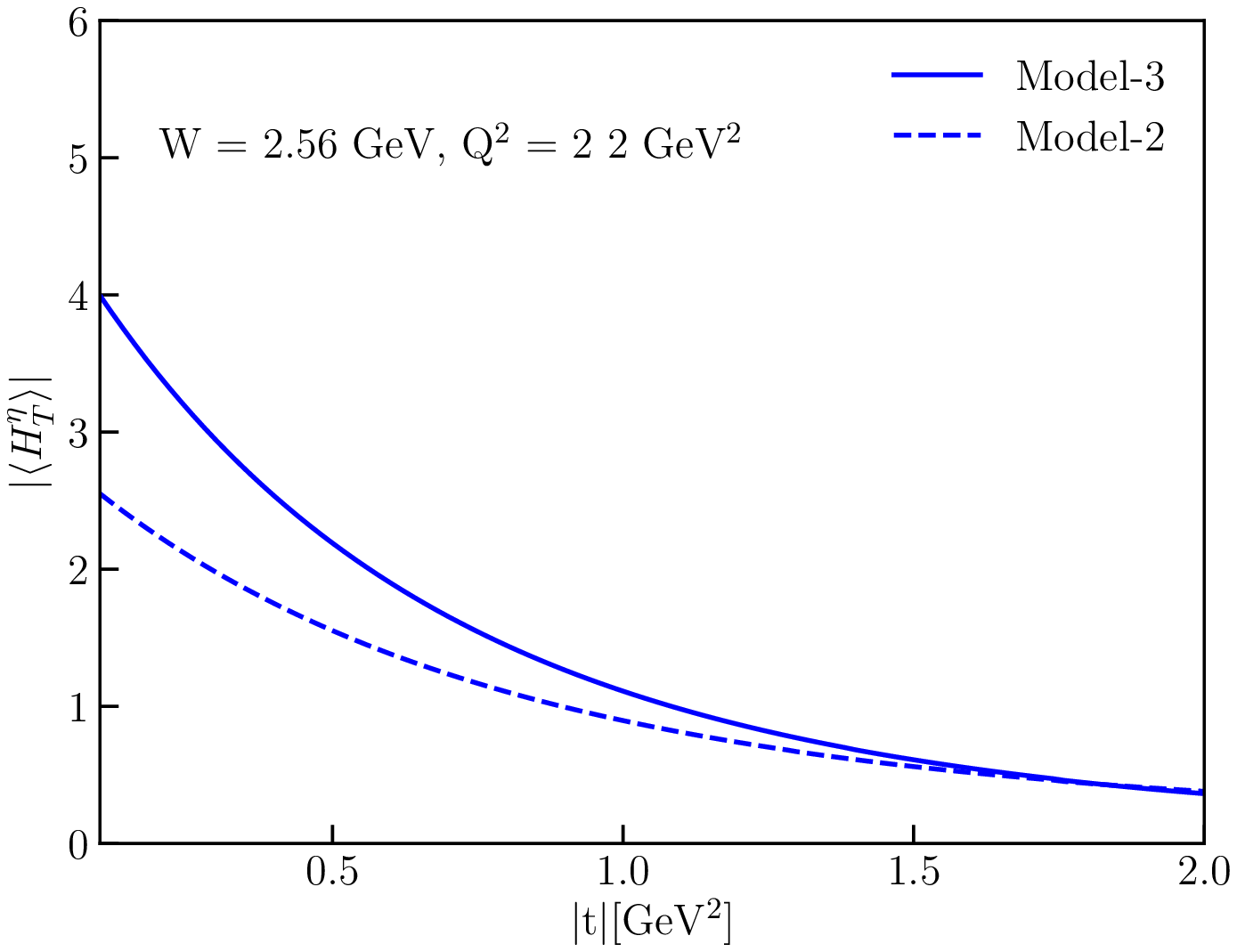}&
            \includegraphics[width=7.5cm]{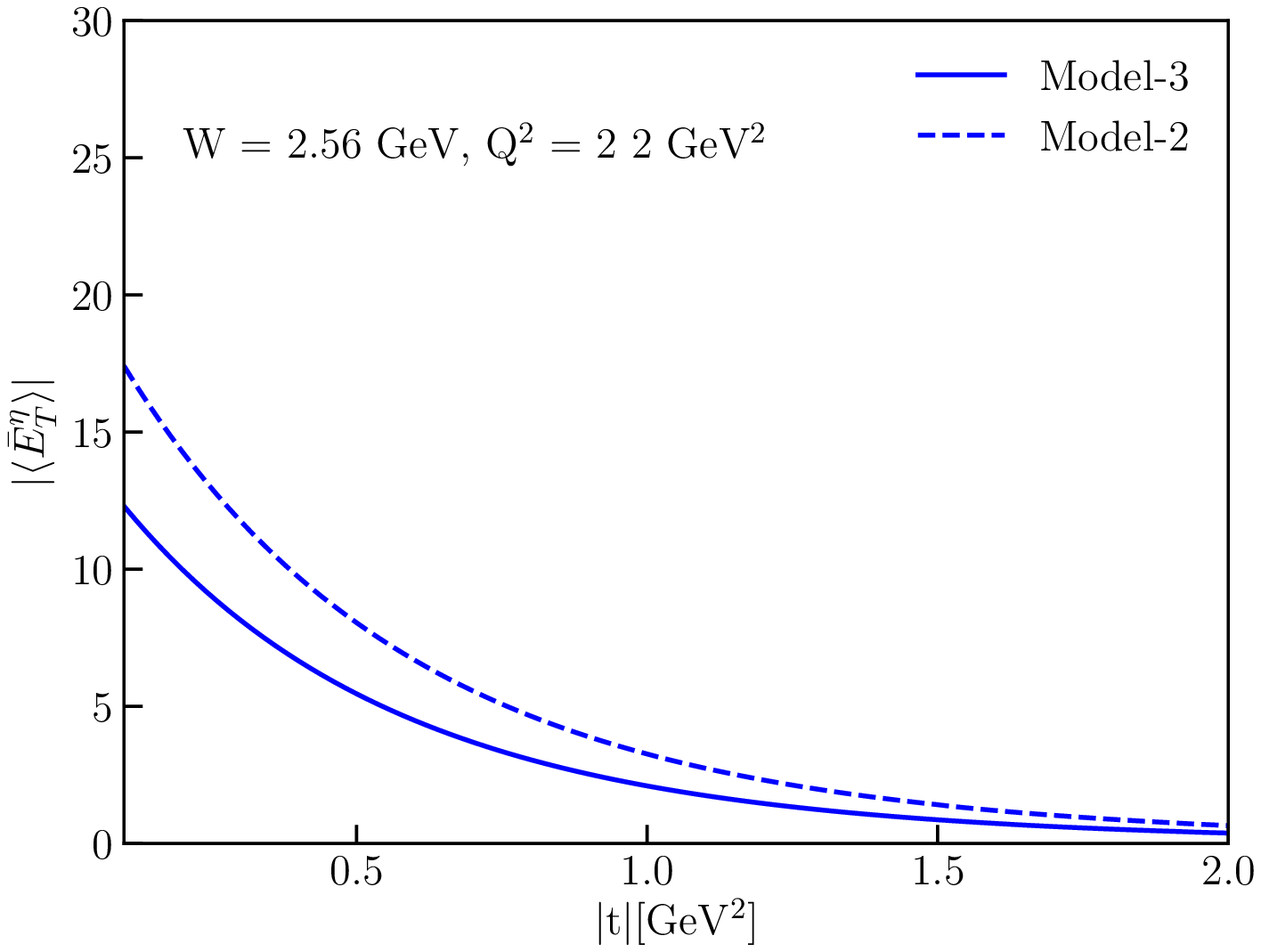}
        \end{tabular}
    \end{center}
    \caption{Extracted from the cross section transversity convolutions $|\langle
        {H_T} \rangle|$ and $|\langle {\bar E_T} \rangle|$ for $\pi^0$
        (upper part) and $\eta$ production (lower part) at CLAS energy
        range.
    }
\end{figure}

Model-2 predictions at EicC energies for $\pi^0$ production are
presented at \cite{gxc22}. For Model-3 at these energies we have
results similar to shown in Fig,~2. The $\pi^0$ cross sections for
Model-3 don't have deep near $ \vert t^\prime\vert =$ 0 $\mbox{GeV}^2$ as we have for Model-2.
Model-2 and 3 results are similar for $ \vert t^\prime \vert \sim$ 0.2 $\mbox{GeV}^2$ and cross
section is a bit smaller for Model-3 with respect to Model-2 at
$\vert t^\prime \vert >$ 0.3 $\mbox{GeV}^2$.

 Our results for EicC energies $W$= 7-16
$\mbox{GeV}$ for $\eta$ production are exhibited in the Figs.~5 and 6. It
can be concluded that Model-3 results are higher for the cross section
$\sigma$ with respect to Model-2 and for $\sigma_{TT}$ result is
opposite- Model-2 gives higher results. This is caused by larger
$H_T$ contribution in Model-3 and larger $\bar{E}_T$ effects in
Model-2 that is important in $\sigma_{TT}$. These model results
can be checked experimentally by EicC and determine what Model-2
or 3 is more adequate to experiment.

\begin{figure}[h!]\label{conv-8}
    \begin{center}
        \begin{tabular}{cc}
            \includegraphics[width=7.5cm]{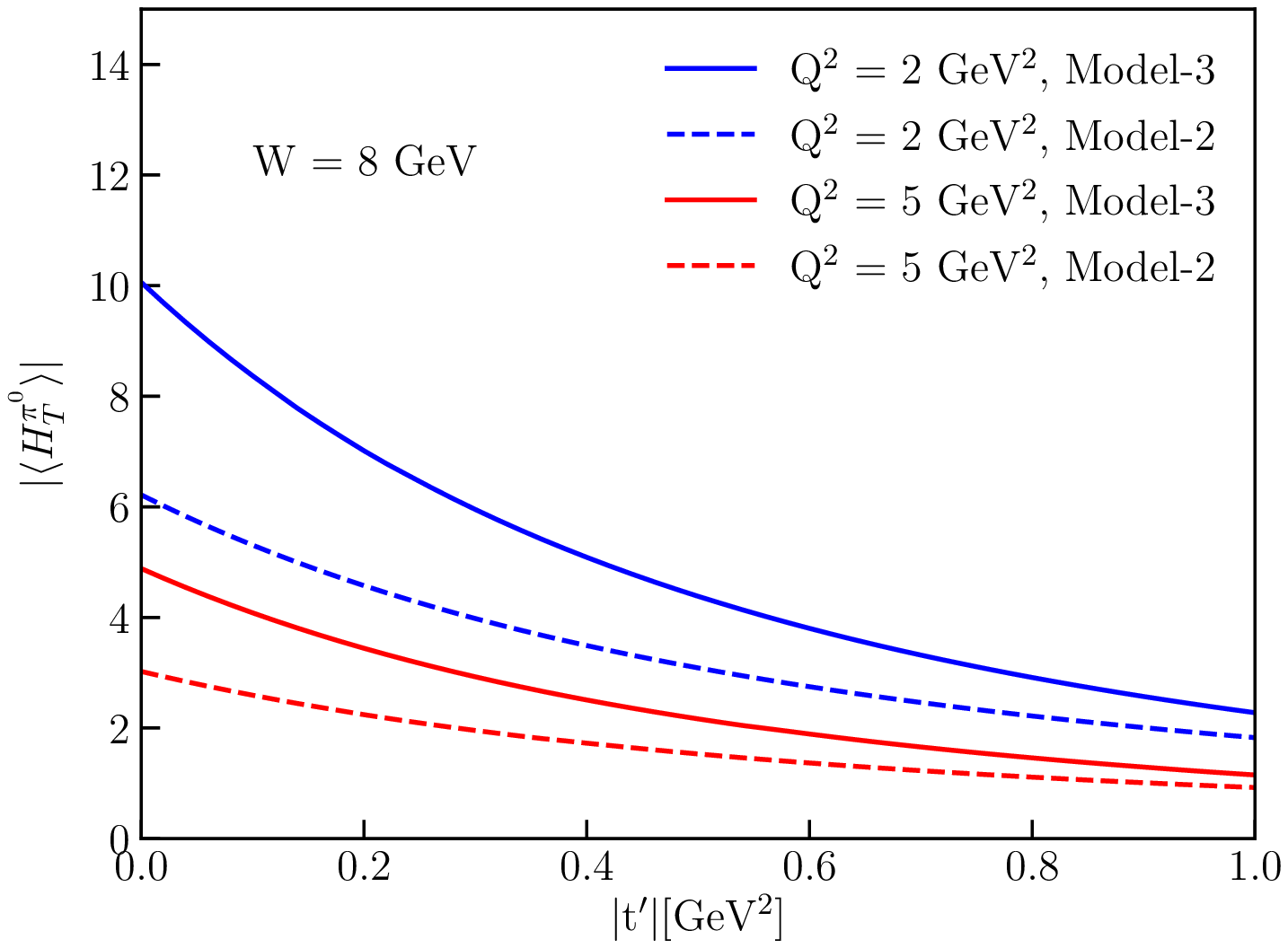}&
            \includegraphics[width=7.5cm]{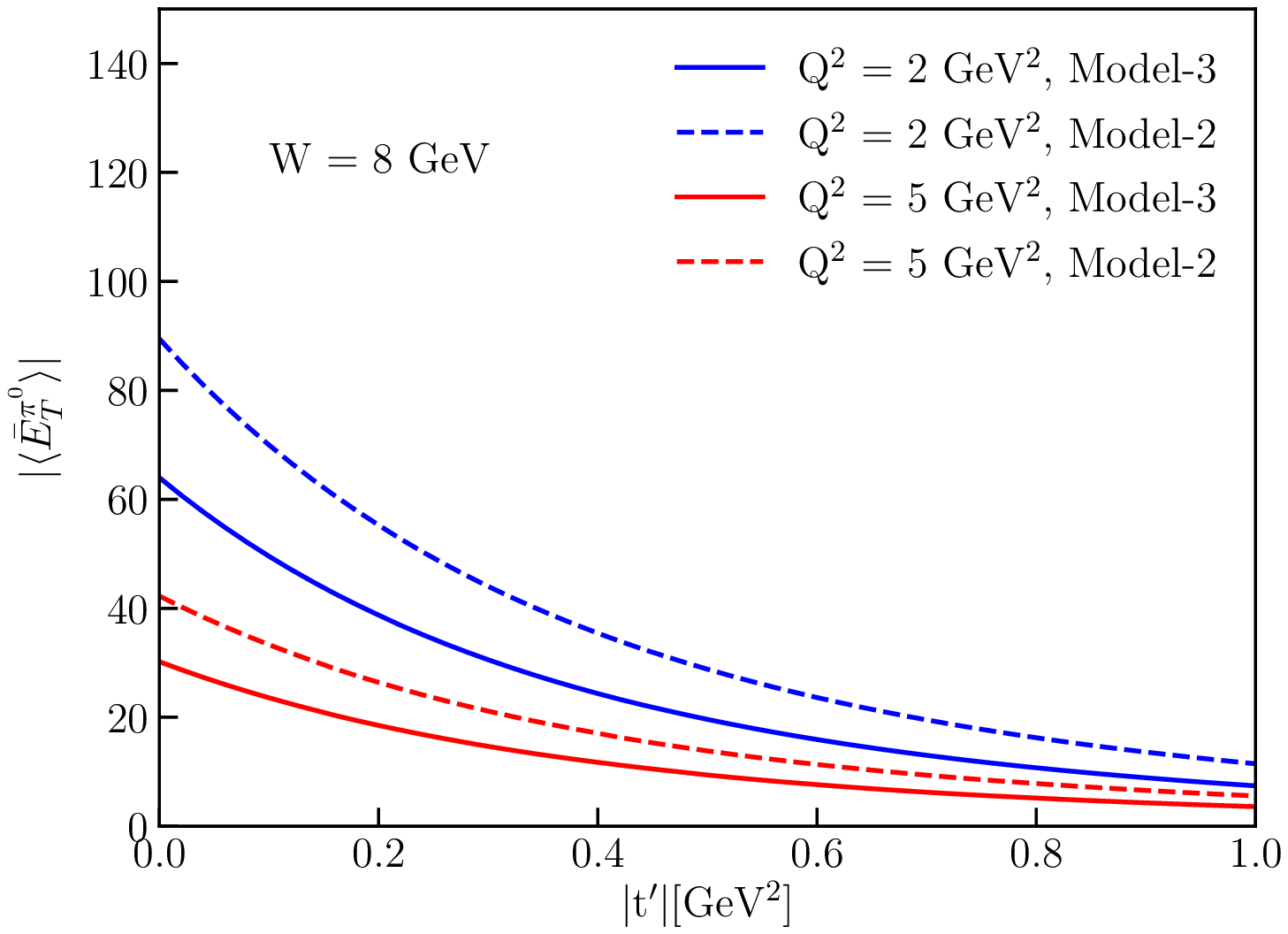}\\
            \includegraphics[width=7.5cm]{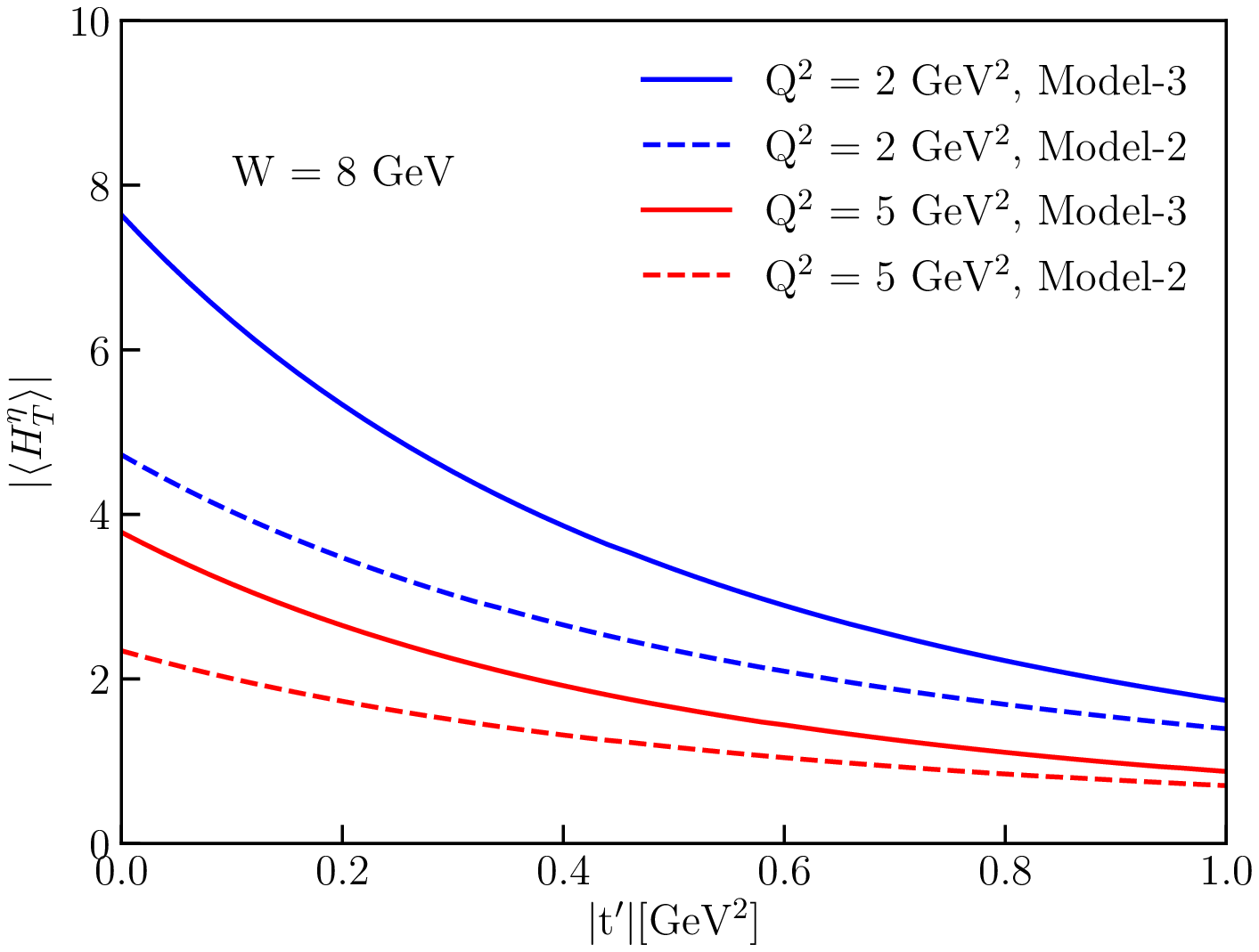}&
            \includegraphics[width=7.5cm]{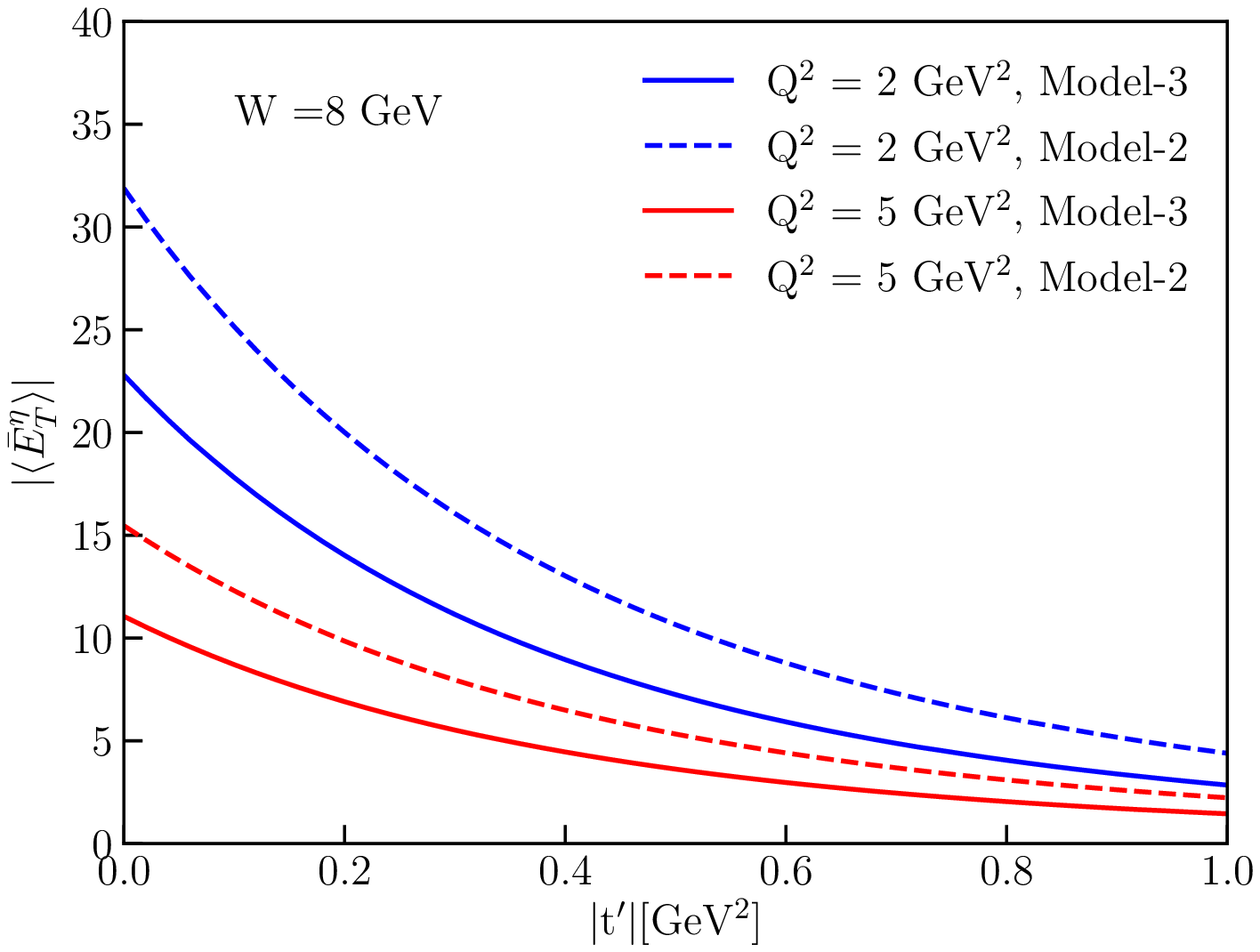}
        \end{tabular}
    \end{center}
    \caption{Extracted from the cross section transversity convolutions for $\pi^0$
    	 (upper part)
        and $\eta$ (lower part) production at EicC ($W$ = 8 $\mbox{GeV}$).
    }
\end{figure}

\begin{figure}[h!]\label{conv-12}
    \begin{center}
        \begin{tabular}{cc}
            \includegraphics[width=7.5cm]{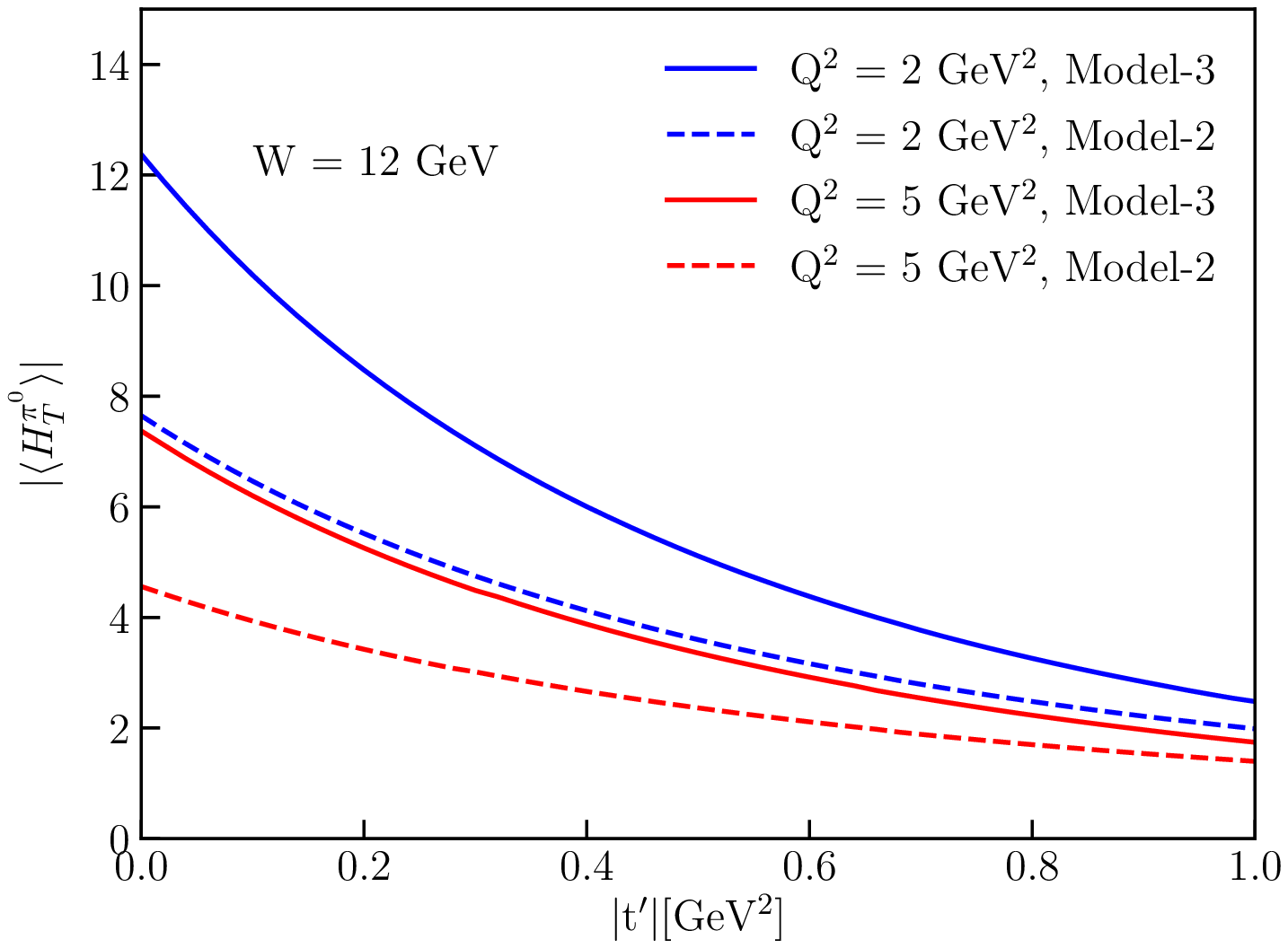}&
            \includegraphics[width=7.5cm]{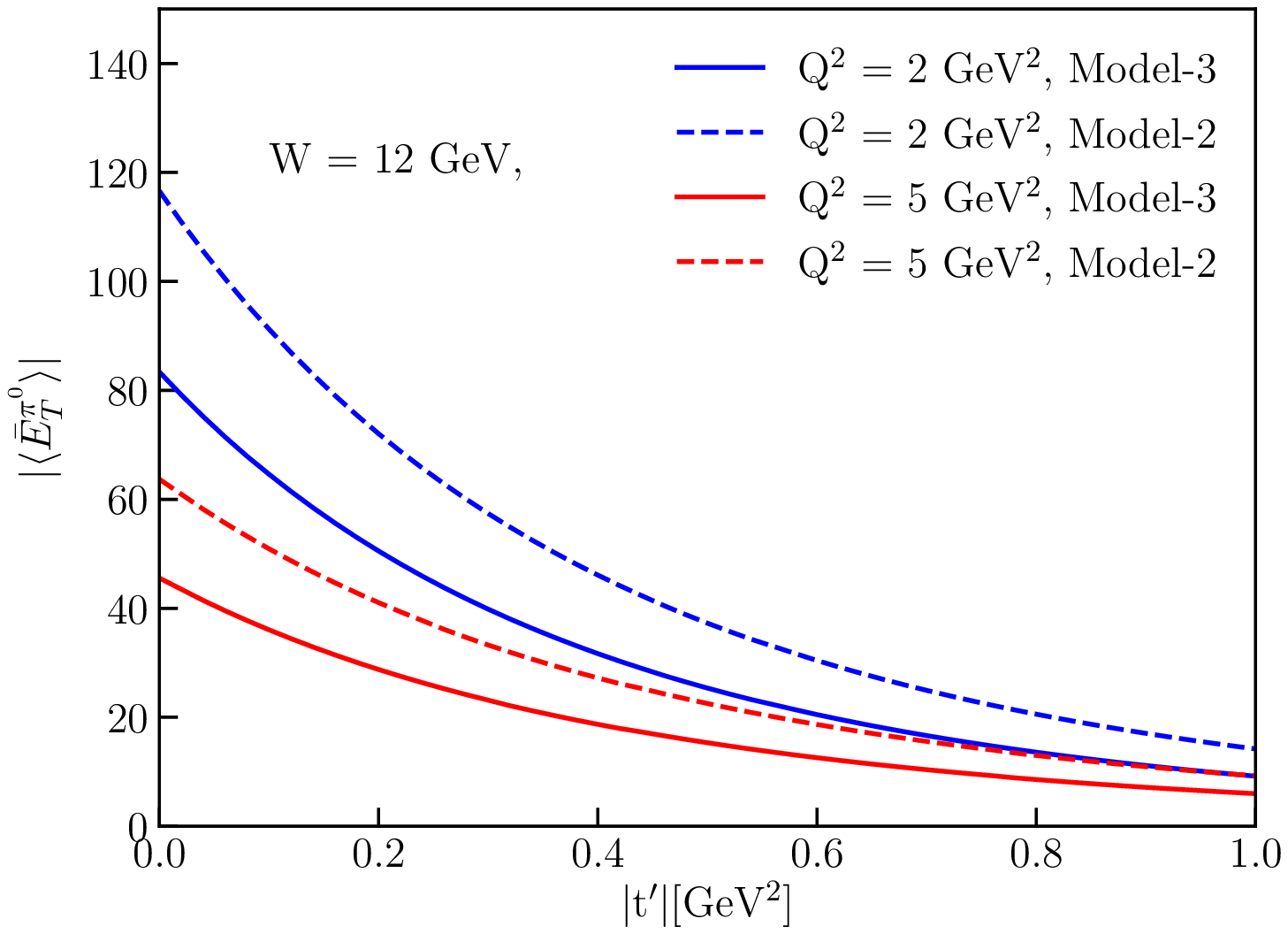}\\
            \includegraphics[width=7.5cm]{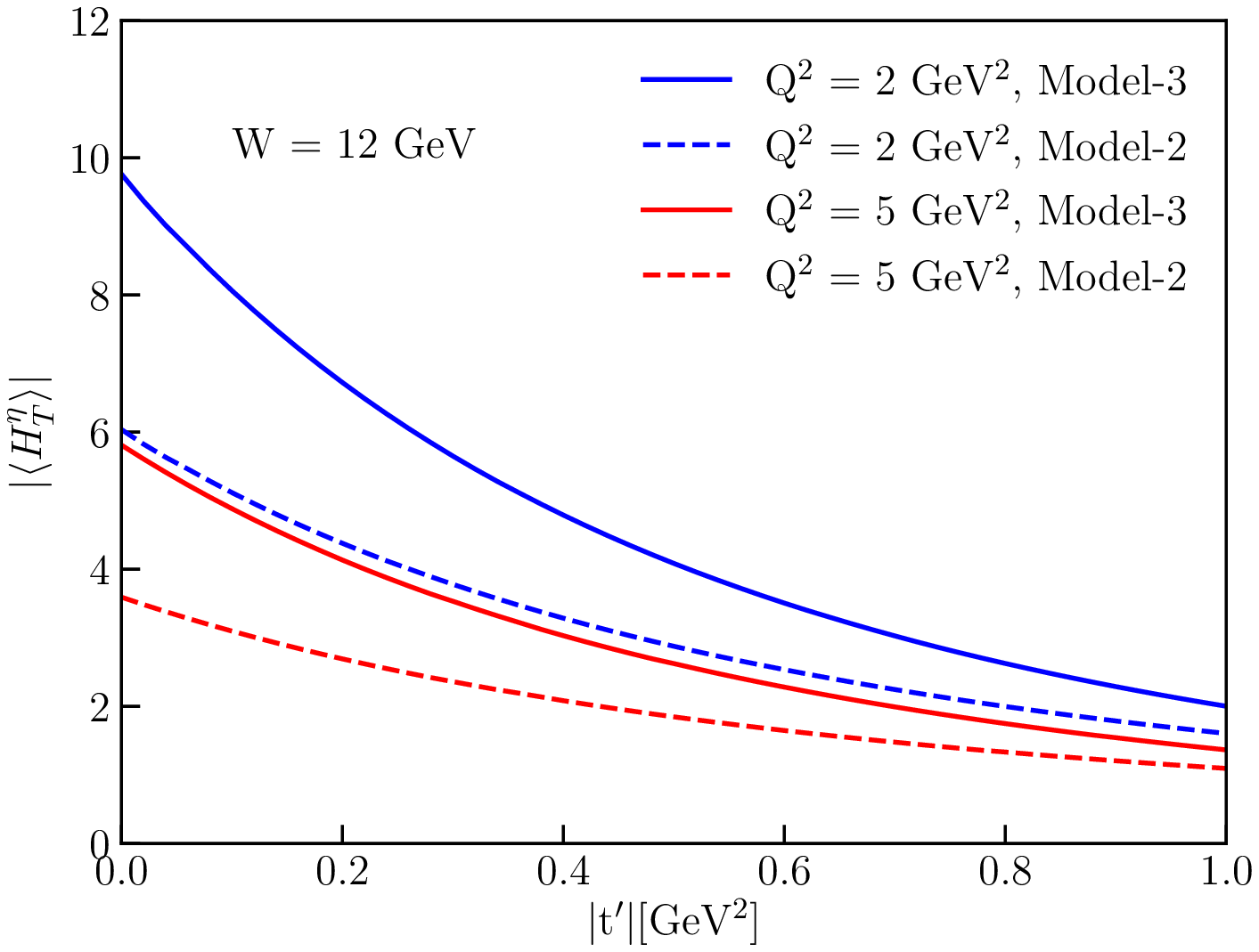}&
            \includegraphics[width=7.5cm]{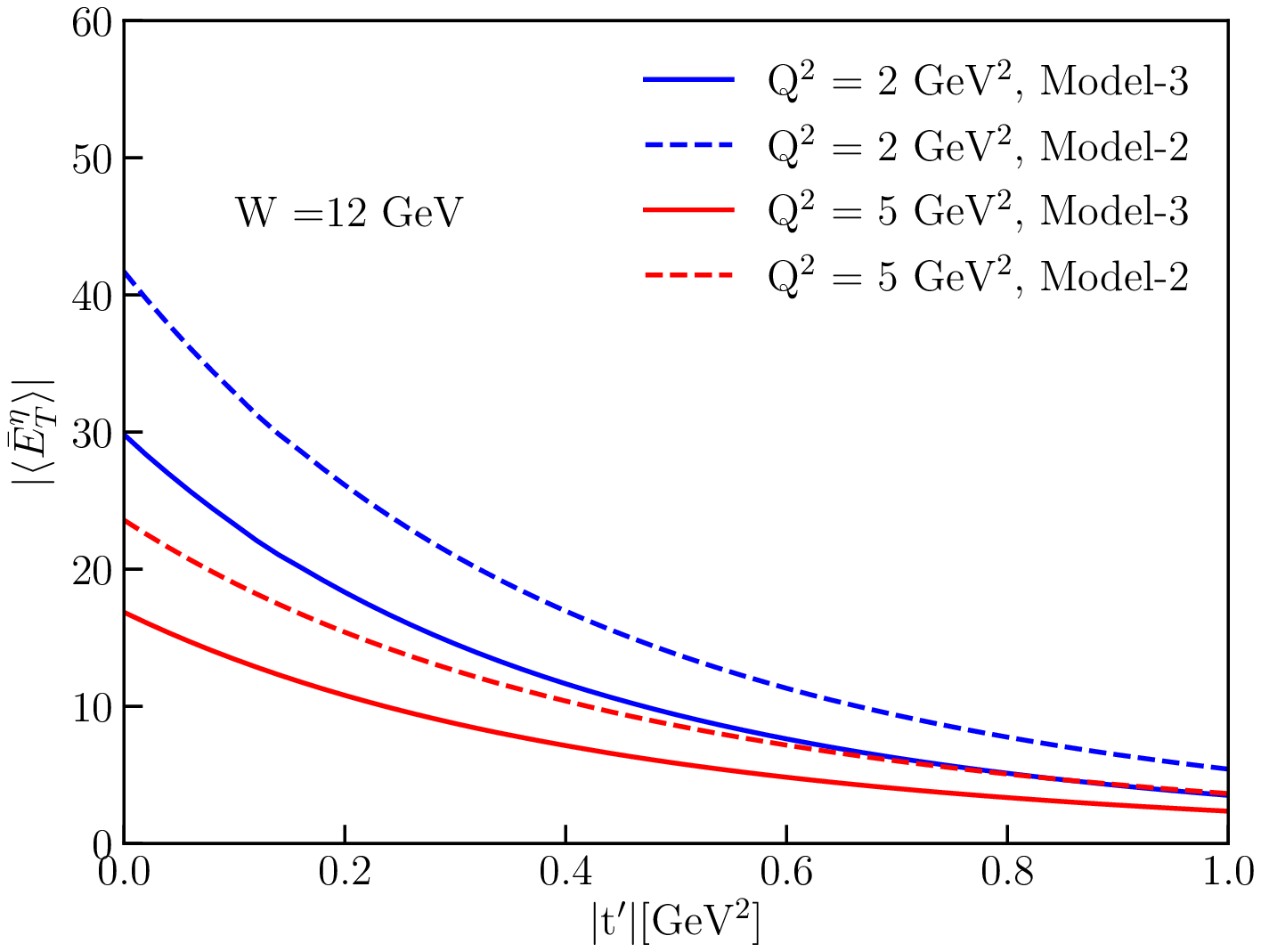}
        \end{tabular}
    \end{center}
    \caption{Extracted from the cross section transversity convolution functions for $\pi^0$ (upper part)
        and $\eta$ (lower part) production at EicC ($W$ = 12 $\mbox{GeV}$).
    }
\end{figure}
Now we shall discuss how we can get information about transversity convolutions
$\bar E_T$ and $H_T$ from experimental data. From Eq.~(\ref{ds}),
 we can obtain
\begin{eqnarray}\label{ampl}
|M_{0+++}|&=&\sqrt{-\kappa \frac{d\sigma_{TT}}{dt}},\nonumber\\
|M_{0-++}|&=&\sqrt{2 \kappa
    (\frac{d\sigma_{T}}{dt}+\frac{d\sigma_{TT}}{dt})},
\end{eqnarray}
we can determine the absolute values of the amplitudes. Employing
normalization factor from Eq.~(\ref{conv}) we can determine $H_T$ and
$\bar E_T$ convolutions. This procedure was adopted to extract tranvsersity
convolutions from CLAS experimental data in Ref. \cite{kr-conv,kubar}.

Now we don't have experimental data from China EicC. To
demonstrate what can be done we shall use instead realistic
experimental data, our model calculations for the cross sections
$\frac{d\sigma_{T}}{dt}$ and $\frac{d\sigma_{TT}}{dt}$. Our
results for  $H_T$ and $\bar E_T$ convolutions for $\pi^0$ and
$\eta$ production are depicted in Fig.~7. They are close to results
found in \cite{kr-conv,kubar} at CLAS energies. As expected we
find that $H_T$ convolution is larger for Model-3, at the same for
Model-2 we get larger $\bar E_T$. Using these results, we can
extract convolutions for $u$ and $d$ flavors under the help of:
\begin{eqnarray}\label{flav}
F^u=\frac{3}{4}(\sqrt{2} F^{\pi}+\sqrt{6} F^{\eta}),\nonumber\\
F^d=\frac{3}{2}(\sqrt{2} F^{\pi}-\sqrt{6} F^{\eta}),
\end{eqnarray}
which is a consequence of Eq.~(\ref{flf}). Here $F$ are
corresponding transversity $H_T$ or $\bar{E}_T$ convolution
functions. Such analyzes was performed at CLAS energies at
\cite{kubar}.

We will not do this here, because we have model results for flavor
convolutions, but extraction of transversity convolution functions from
future experimental data can be in important result in later experiments.

Our predictions for $H_T$ or $\bar {E}_T$ convolution functions that were
extracted from the cross sections at the energies $W=$ 8, 12 $
\mbox{GeV}$ which are typical at EicC energy range are exhibited in
Figs.~8 and 9. Experimental analyses of these quantities can give
information on the preferable models for transversity GPDs.

In Fig.~10, we present our model predictions for energy dependencies of
transversity convolution functions at fixed $Q^2$ and momentum transfer.
Such analyses will be important to give a constraints on the $W$-
dependence of $H_T$ or $\bar {E}_T$ GPDs from future experimental
data.

Note that the transversity dominance $\sigma_T\gg \sigma_L$ that
was tested for $\pi^0$ production at high energies is valid for
$\eta$ production at the energies $W = $2$\sim$ 15 $\mbox{GeV}$. This
means that in experimental analyzes of transversity convolutions,
unseparated cross section $\sigma = \sigma_T+ \epsilon \sigma_L$ can
be applied instead $\sigma_T$ for both processes of $\pi^0$ and
$\eta$ production.

\begin{figure}[h!]\label{convW}
    \begin{center}
        \begin{tabular}{cc}
            \includegraphics[width=7.5cm]{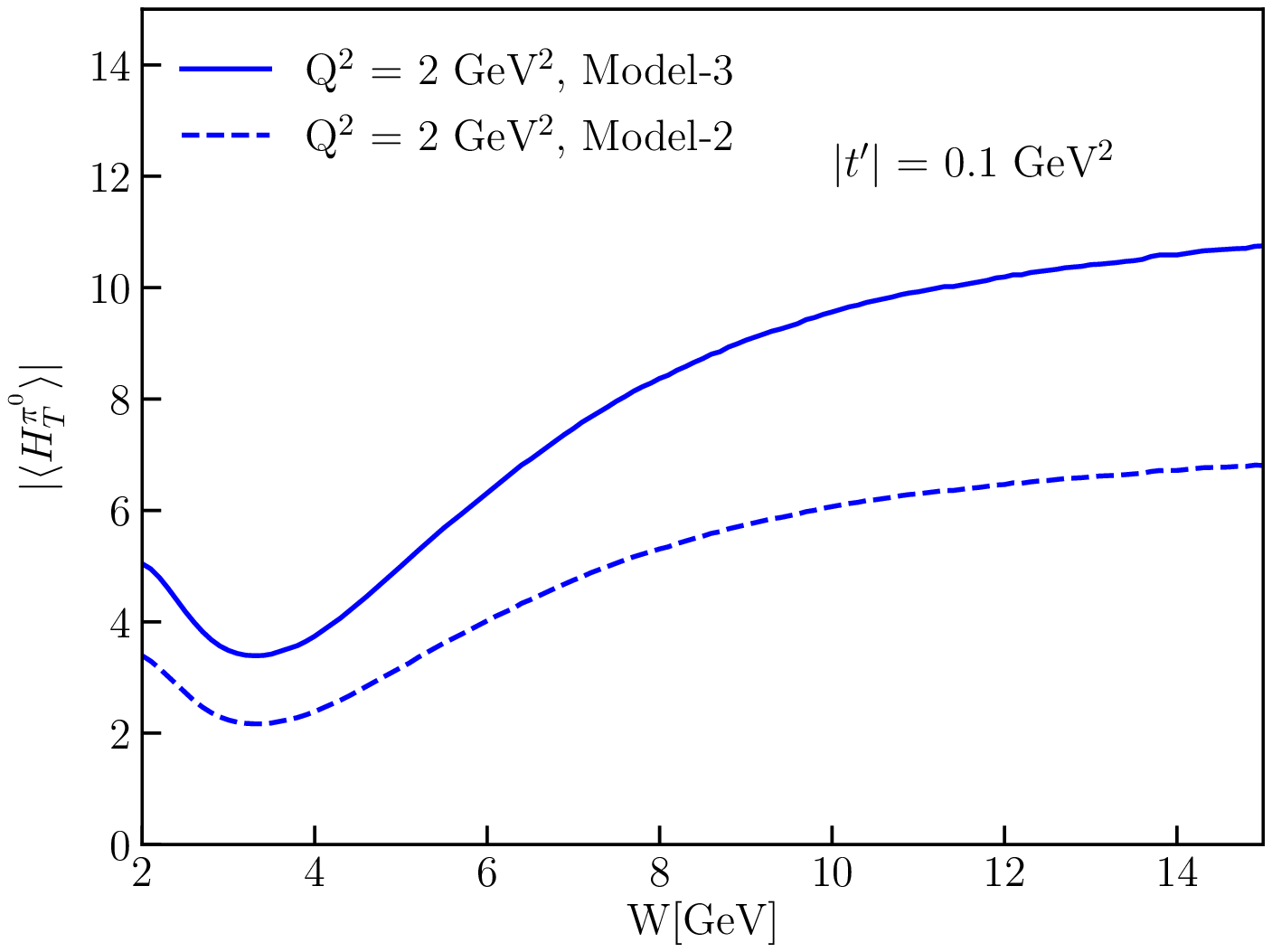}&
            \includegraphics[width=7.5cm]{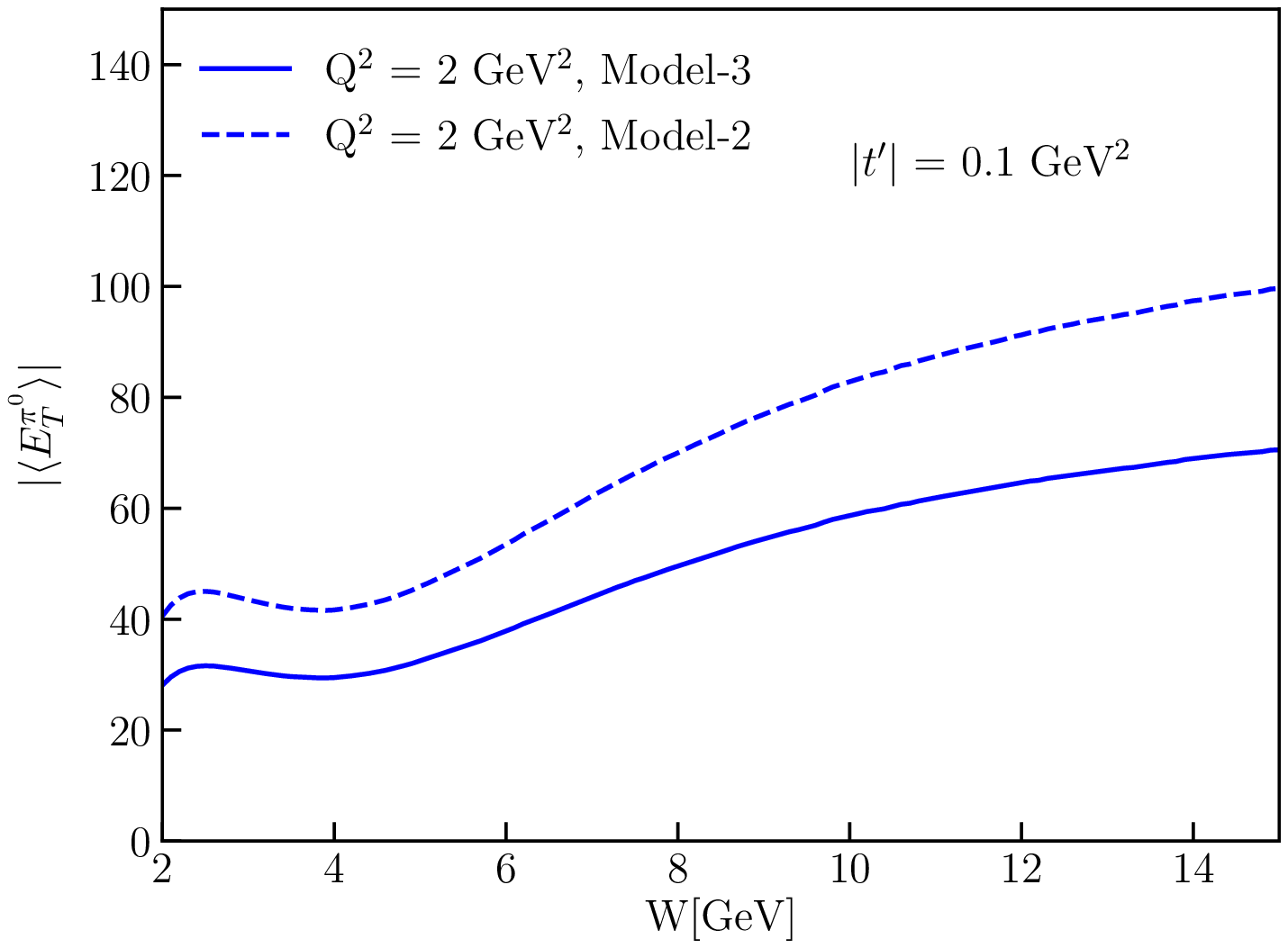}\\
            \includegraphics[width=7.5cm]{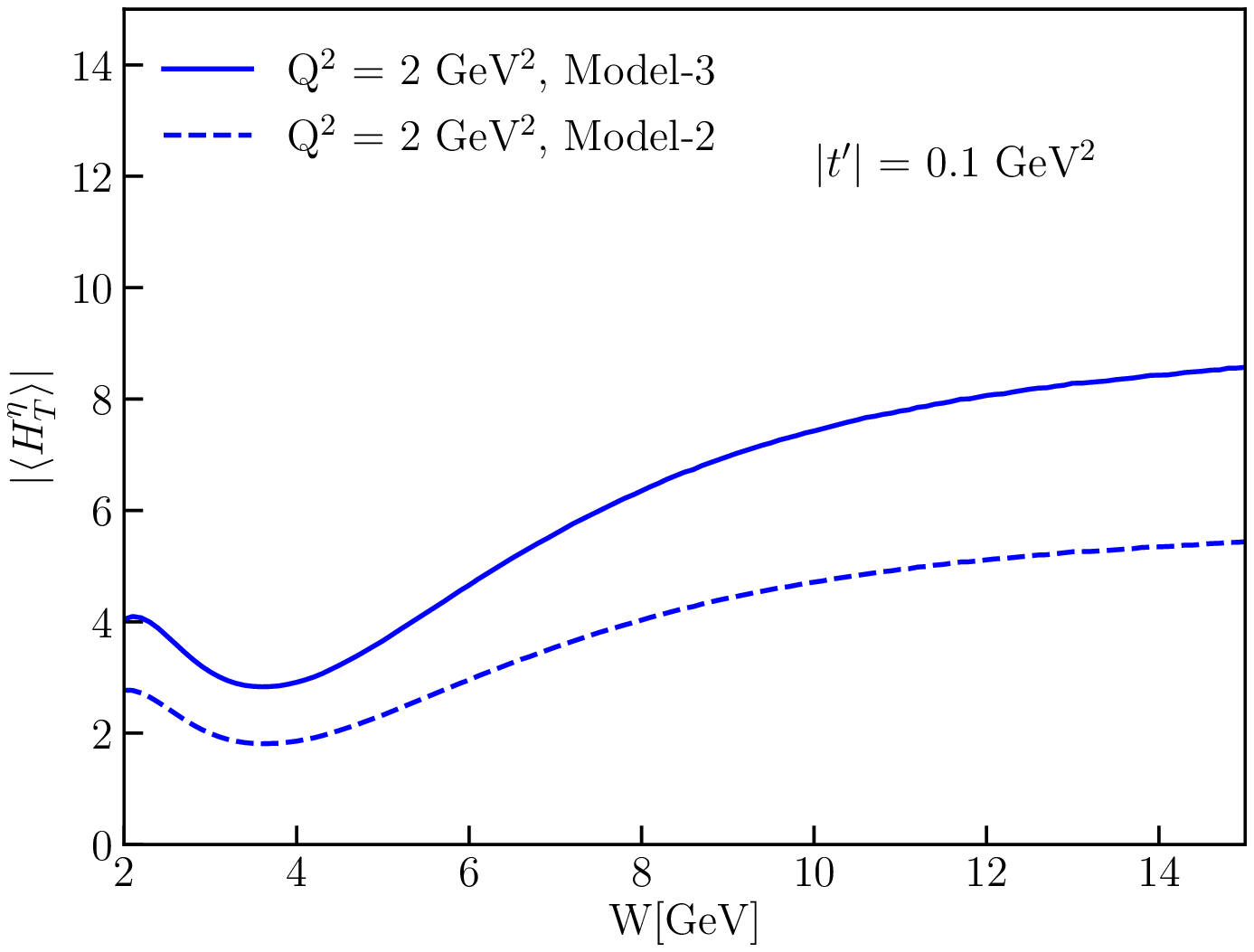}&
            \includegraphics[width=7.5cm]{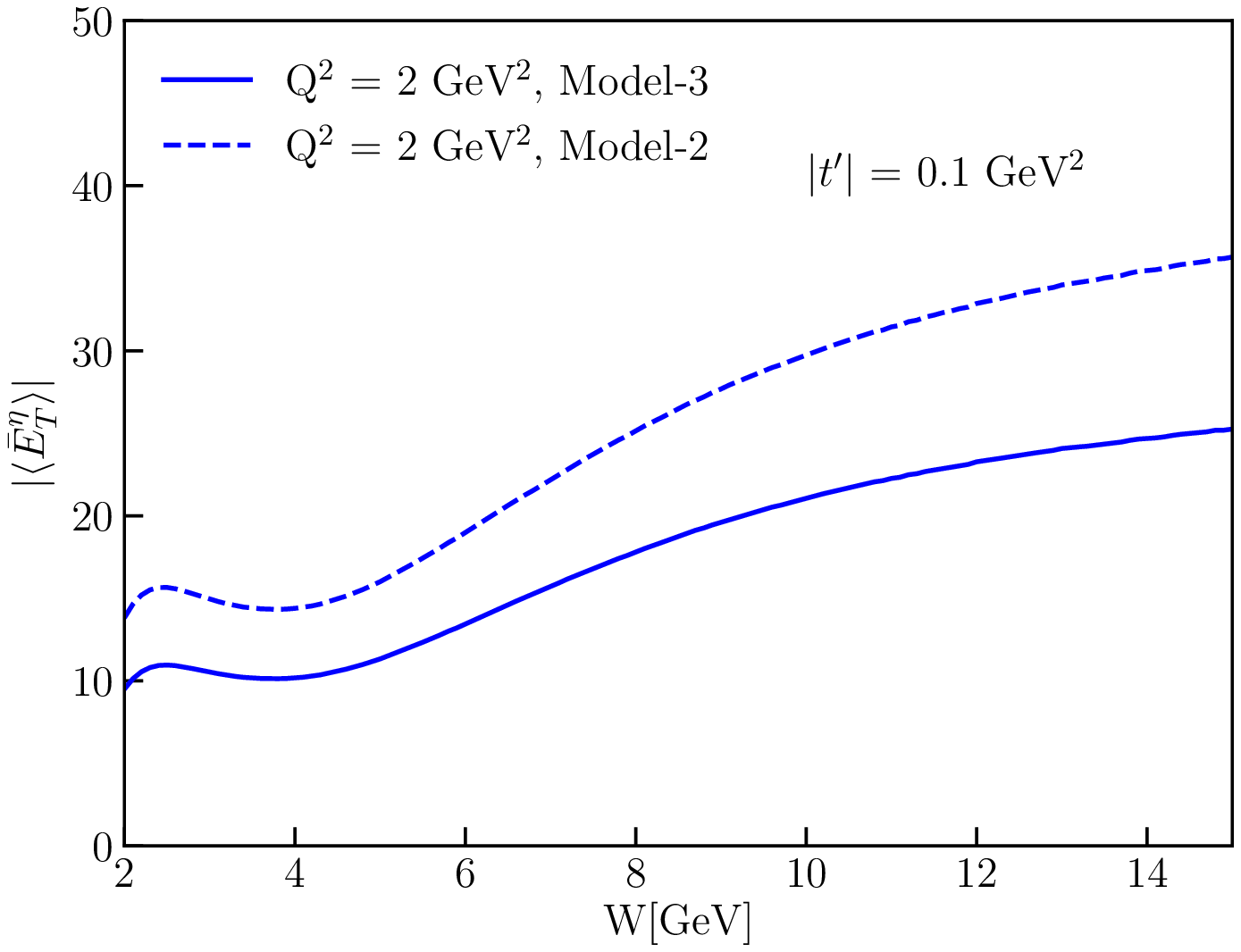}
        \end{tabular}
    \end{center}
    \caption{Energy dependencies of extracted trasnversity convolutions for $\pi^0$ (upper part)
        and $\eta$ (lower part) production at EicC ($Q^2$ = 2 $\mbox{GeV}^2$, $|t'|$ = 0.1 $\mbox{GeV}^2$).
    }
\end{figure}

\section{Conclusion}
In this paper we investigate exclusive electroproduction of
pseudoscalar $\pi^0$ and $\eta$ meson at China EicC energies. The
process amplitudes are calculated in the model where amplitudes
are factorized into subprocess amplitudes and GPDs. For the
transversity twist-3 effects, the subprocess amplitude ${\cal
    H}_{0-,++}$ is the same in Eq.~(\ref{ht}) for both contributions that
contains $H_T$ and $\bar{E}_T$ GPDs.

We consider two GPDs parameterization Model-2 and  Model-3. Both
models describe properly $\pi^0$ and $\eta$ production at CLAS
energies. It seems that Model-3 gives better results for $\pi^0$
production at COMPASS at large momentum transfer and gives better
description of $\eta$ production at CLAS at momentum transfer $|t|
< $0.5 $ \mbox{GeV}^2$.

We perform predictions for unseparated $\sigma$ and $\sigma_{TT}$
cross sections for EicC kinematics for $\eta$ production with
Model-2 and Model-3. We observe that transversity dominance $\sigma_T\gg
\sigma_L$, found at low CLAS energies \cite{gk11} and confirmed at
EicC energies in Ref.\cite{gxc22} for $\pi^0$ process is valid at all
these energies for $\eta$ production too.

Adopting combination of the cross sections, we extract GPDs $H_T$ and
$\bar E_T$ convolutions determined in Eq.~(\ref{conv}) for the cases
of $\pi^0$ and $\eta$ mesons. These results for EicC energies are
quite different that give possibility to determine preferable
model at future experiments. In addition, we analyze energy
dependencies of transversity convolutions at fixed $t$ and $W$
that can give information about energy parameters of GPDs $H_T$
and $\bar {E}_T$ from the data.

Note that reactions $\pi^0$ and $\eta$ production considered
here have different flavor contributions to the amplitudes. This
gives possibility to perform $u$ and $d$ flavor separation for
transversity GPDs \cite{kubar}.

Our results can be useful in future experiments at China EicC on
the pseudoscalar mesons production and give more important
knowledge on transversity influences at these energy ranges.

\section*{Acknowledgment}
S.G. expresses his gratitude to P.Kroll for collaboration on GPDs
study and to V.Kubarovsky for important discussions. The work 
partially supported by is Strategic Priority Research Program of 
Chinese Academy of Sciences (Grant
NO. XDB34030301) and the CAS president's international
fellowship initiative (Grant No. 2021VMA0005). 

\end{document}